\documentstyle[aps,prb]{revtex}
\begin{document}

%\twocolumn[
 \draft
\title{ Theory of anyon excitons:\\
 Relation to excitons of $\nu =1/3$ and $\nu=2/3$ incompressible liquids}
\author{}
\address{{\rm M. E. Portnoi}\\
Department of Physics, University of Utah, Salt Lake City, UT 84112\\
 and A. F. Ioffe Institute, St. Petersburg 194021, Russia\\
 {\rm E. I. Rashba}\\
Department of Physics, University of Utah, Salt Lake City, UT 84112\\
and L. D. Landau Institute for Theoretical Physics, Moscow 117940, Russia}

\date{Published in Phys. Rev. B {\bf 54}, 13791 (1996)}

\maketitle

%\widetext

\begin{abstract}
 Elementary excitations of incompressible quantum liquids (IQL's)
 are anyons, i.e., quasiparticles carrying fractional charges
 and obeying fractional
 statistics. To find out how the properties of these exotic quasiparticles
  manifest themselves
 in the optical spectra, we have developed the anyon exciton model (AEM)
 and compared the results with the finite-size data for excitons of
 $\nu =1/3$ and $\nu =2/3$ IQL's. The model  considers an exciton
 as a neutral composite consisting of three quasielectrons and a single hole. 
 The AEM works well when the separation between electron and hole confinement
 planes, $h$, obeys the condition $h\agt 2l$, where $l$ is the magnetic
 length. In the framework of the AEM an exciton possesses momentum
 $\bf k$ and two internal quantum numbers, one of which can be chosen
 as the angular momentum, $L$, of the ${\bf k}=0$ state. Charge
 fractionalization manifests itself in striking differences between the
 properties of anyon excitons and ordinary magnetoexcitons. Existence
 of the internal degrees of freedom results in the multiple branch energy
 spectrum, crater-like electron density shape and 120$^{\circ}$ density
 correlations for ${\bf k}=0$ excitons, and the splitting of the electron
 shell into bunches for ${\bf k}\neq 0$ excitons. For $h\agt 2l$ the bottom
 states obey the superselection rule $L=3m$, $m\geq 2$ are integers, and
 all of them are hard core states. For $h\approx 2l$ there is one-to-one
 correspondence between the low-energy spectra found for the AEM and
 the many-electron exciton spectra of the $\nu =2/3$ IQL, whereas some
 states are absent from the many-electron spectra of the $\nu =1/3$ IQL.
 We argue that this striking difference in the spectra originates
 from the different populational statistics of the quasielectrons of
 charge conjugate IQL's and show that the proper account of the
 statistical requirements eliminates excessive states from the
 spectrum. Apparently, this phenomenon is the first manifestation
 of the exclusion statistics in the anyon bound states.
\end{abstract}


%\narrowtext

\section{Introduction}
\label{sec:introduction}

 Incompressible quantum liquids (IQL's)\cite{Laug83} underlie the
 fractional quantum Hall effect (FQHE) discovered by Tsui, Stormer, and
 Gossard.\cite{TSG82} The charge carriers in these liquids are
 anyons, i.e., quasiparticles (quasielectrons and quasiholes)
carrying fractional charges\cite{Laug83} and obeying fractional
 statistics.\cite{Halp84,ASW84}
 Historically the main experimental discoveries in this
 field were done by magnetotransport experiments. However, the
 role of spectroscopic methods is continuously increasing since
 they provide an indispensable tool for investigating spectra of
 elementary excitations. Fine structure 
 specific for different
 electronic phases was discovered in the spectra of radiative
 photoemission.\cite{emission} These spectra were used to measure
 gaps in the energy spectra of IQL's, for investigation of
  phase transitions between the IQL and Wigner solid phases, etc.
 The frequency of long-wave neutral elementary excitations of IQL's
 was measured in Raman scattering experiments.\cite{PinDPW93}

 A challenging problem in physics of IQL's is a direct observation
 of the charge fractionalization. Between different exciting approaches
 to this problem the spectroscopic approach seems to be one of the most
 promising. Indeed, spectroscopy permits one to observe properties of
 IQL's in the bulk where the effect of the impurities and edges is 
 reduced to the minimum. The  intrinsic spectroscopy of
 IQL's is the magnetospectroscopy of excitons. However,  properties
 of excitons  reflect the spectrum of the elementary excitations of an
 IQL (quasielectrons and quasiholes, magnetorotons,\cite{GMP86} etc.) and
 can be treated in terms of them only when the separation $h$ between
 electron and hole confinement planes is sufficiently large. Indeed,
 when $h\alt l$, where $l$ is the magnetic length, the filling factor
 $\nu$ of the liquid strongly deviates in the vicinity of the hole
 from the quantized value $\nu = p/q$. Under these conditions the
 properties of the IQL cannot be treated in terms of its quasiparticles.
 The spectroscopy of a remote hole has been discussed from different
 standpoints in a number of papers.\cite{MRK,Birm92,ARPis92,RP93,CQ94}
 Despite the fact that experiments in the $h\gg l$ region are rather
 complicated, the separations up to $h\approx 5l$ were achieved in experiments
 on extrinsic radiative photoemission.\cite{hdepend94} Of special
 importance might be experiments performed for a fixed filling factor
 $\nu=p/q$ and a variable dimensionless electron-hole separation
 $h/l$.\cite{RA} The first experiments of this kind were reported
 recently.\cite{Heiman96}

 The investigation of excitons is also important from the different
 standpoint. In the theory of IQL's  the
 statistical properties of the system of free anyons are usually
 discussed. In an exciton the
 anyons exist in a bound state because of the attracting potential
 of a hole. It was shown\cite{APR,AR2/3}
 that the energy spectra of excitons
 of $\nu = 1/3$ and $\nu = 2/3$ IQL's  are closely related to the 
 statistical and dynamical properties of quasielectrons of these
 IQL's. Therefore, the exciton problem can be really treated as
 a few-anyon  problem. Excitons represent a wide class 
 of few-anyon systems. It includes excitons, anyon ions\cite{CQ93,Rez93}
 (which can be also treated as quasiparticle-exciton complexes\cite{AR-PR}),
 some impurity centers, etc. In what follows, we consider excitons
 as bound few-anyon composites and investigate their properties
 in some detail.

There are two approaches to the theory of excitons in IQL's.
 The first approach is based on the exact diagonalization for finite-size
 systems in the spherical geometry.\cite{Hal83} Having in mind
 workable system sizes, it provides reliable results for $h\alt 2l$,
 at least as applied to the $\nu = 1/3$ and $\nu = 2/3$ IQL's. The
 second approach is based on the anyon exciton model (AEM)
 proposed by the present
 authors.\cite{RP93,MPL} In the framework of this approach an exciton is
 considered as a neutral composite quasiparticle consisting of several
 anyons and a hole. This model is exact only when the exciton size which
 is about $h$ is large  compared to the size of anyons which is
 about $l$, i.e., for $h/l \gg 1$. Therefore, the two approaches are
 complementary and one can expect that they match when $h/l \sim 1$. 

 It is the main statement of the AEM that excitons of IQL's
 possess a multiple branch spectrum. Indeed, a charged particle at the
 lowest Landau level possesses a single degree of freedom and a single
 quantum number. An exciton being a neutral entity possesses in a 
 magnetic field a vector momentum $\bf k$ absorbing two degrees of
 freedom.\cite{GD}  Therefore, an exciton consisting of $q$ anyons and
 a hole possesses $q-1$ internal degrees of freedom. For an ordinary
 magnetoexciton $q=1$ and the spectrum consists of a single branch.
 For $q\geq 2$ an exciton acquires internal quantum number(s) and
 multiple branch spectrum. This prediction of the AEM permitted
 Apalkov {\it et al.}\cite{APR,AR2/3,ARCamb} to represent the energy
 spectra found by finite-size computations for $\nu= 1/3$ and $\nu =2/3$
 IQL's as a system of exciton branches and to assign to these branches
 internal quantum numbers. The latter determine the values of the
 exciton angular momentum, $L$, in the ${\bf k}=0$ states.
 Zang and Birman\cite{ZB95} and Chen and Quinn\cite{CQ95} also inferred
 existence of several exciton branches in their finite-size data. 

 In the range accessible for finite-size studies, $h\alt 2l$, there are
 two types of excitons in the lower part of the energy spectrum,
 anyon excitons and tight excitons. 
 {\it Anyon excitons} are loose entities with a pronounced
 anyon-density dip  at the center.
  They are generically related to the quantum states making up 
 the low-energy (anyon) sector\cite{HXZ,DJ92} of the electron subsystem.
  Each anyon exciton is a bound state of a three-quasielectron complex
 from the anyon sector and a hole. This finding establishes a connection
 between the spectroscopy of excitons and the low-energy physics of
 the FQHE. There is a striking difference in the quantum numbers of the
 low-energy anyon excitons of the $\nu =1/3$ and $\nu =2/3$ IQL's. 
 It is related to the difference in
 the energy spectra of the three-quasielectron complexes
 originating due to the different populational statistics
 of the quasielectrons of these liquids. {\it Tight
 excitons} are dense entities. A sharp density maximum is achieved either
 in the center of an exciton or in a close vicinity of the center.
 These excitons are not related to the low-energy sector of the electron
 subsystem. The $L=0$ tight exciton is the bottom state of the exciton
 spectrum in all the region $h\alt 2l$. Therefore, the $h\agt 2l$
 region seems to be of the most importance for the study of the anyon
  substructure of excitons and for the spectroscopic observation
 of the charge fractionalization.

 We suppose everywhere in this paper that the spin-polarized background
 is stable with respect to formation of charged spin-textures
 (skyrmions). It was predicted recently\cite{CoopCh} that spin-textures
 develop in the ground state of a $\nu = 1$ exciton for large values
 of $h/l$; the critical value of $h/l$ is about 1.5 for GaAs.
 Nevertheless, we feel that the spin-polarized ground state of a
 $\nu = 1/3$ magnetoexciton is stable because skyrmions can exist only
 at very low magnetic fields when $\nu \approx 1/3$.\cite{KWJ96}
 Spin-depolarized excitons should be seen in the spectrum of
 excited states.

 In this paper we develop the theory of excitons in the framework
 of the AEM as applied to excitons consisting of three quasielectrons
 and a hole. We use Halperin
 pseudo-wave-functions\cite{Halp84}  in a boson basis and
 investigate the energy spectrum, electron density distribution,
 and anyon correlations in an exciton. We make comparison with the
 finite-size data\cite{APR,AR2/3} 
 on the excitons of the $\nu =1/3$ and $\nu =2/3$  IQL's and conclude that
 excitons of the $\nu = 2/3$ IQL are described rather well by the AEM
 because of the bosonic populational statistics 
 and the narrow form-factors  of the quasielectrons of this liquid.
  We also relate the difference in the energy
 spectra of the anyon excitons of the $\nu =1/3$ and $\nu =2/3$ IQL's
 to some specific features in the exciton shape found in the
 framework of the AEM. For both IQL's the bottom exciton states
 are made by tight $L=0$ excitons for $h\alt 2l$ and by a succession
 of anyon excitons for  $h\agt 2l$. This succession
 consists of hard core excitons with the angular momenta $L$
 which are integers of 3 and increase as $L\propto h^2$ with $h$.

 The general outline of the paper is as follows. In Sec.~\ref{sec:functions}
 we construct a full basis of the exciton wave functions for the AEM
 using the translational symmetry and permutation symmetry arguments.
 In Sec.~\ref{sec:Hamilt} we develop a technique for calculating
 different matrix elements entering the Schroedinger equation. In
 Secs.~\ref{sec:Energy}, \ref{sec:density}, and \ref{sec:correlation}
 we obtain energy spectra, electron density distributions, and
 the density correlation functions, respectively. The latter functions
 unveil the anyon substructure of excitons both in the finite-size data
 and in the AEM. In Sec.\ref{sec:comparison} we make a comparison of the
 results obtained in the framework of the AEM with 
 finite-size data of Refs.~\onlinecite{APR} and \onlinecite{AR2/3}.
 We propose that a striking difference in the finite-size data for
 the $\nu =1/3$ and $\nu =2/3$ IQL's originates from the difference
 in the populational statistics of the quasielectrons of these liquids.

\section{Wave functions}
\label{sec:functions}

Let us consider an exciton consisting of a valence hole with a charge
$(+e)$ and three QE's with electrical charges $(-e/3)$ and statistical
charges $\alpha$. Such an entity provides the AEM description
 of the anyon excitons of the $\nu =1/3$ and $\nu =2/3$ IQL's.
 For an $\nu = 1/3$ IQL the statistical charge 
 equals $\alpha = - 1/3$, while for
 an $\nu = 2/3$ IQL the statistical charge has the same value,
 $\alpha = 1/3$, as for quasiholes in an $\nu = 1/3$ IQL. \cite{Halp84}
 In comparison, $\alpha=0$ for bosons and $\alpha=1$ for fermions.
 In the strong magnetic field limit, when the Coulomb energy
 $\varepsilon_{C} = e^{2}/\epsilon l \ll \hbar \omega_{c}$,
 where $\omega_{c}$ is the cyclotron frequency
  and $\varepsilon $ is the dielectric constant,
 it is convenient to employ dimensionless variables scaled in units
 $\varepsilon_{C}$, $l$ and $e$. We use the symmetric gauge, 
 ${\bf A} = {\hat z}\times {\bf r}/2$, where ${\hat z}$ is a unit vector
 perpendicular to the confinement plane. Instead of the hole, ${\bf r}_{h}$,
 and anyon, ${\bf r}_{i}$, coordinates it is convenient to introduce the
 following two-dimensional (2D) coordinates:
\begin{equation}
{\bf R} = {1\over 2}({\bf r}_{h} + {1\over 3}\sum_{i=1}^{3}{\bf r}_{i})~,~
  {\bbox {\rho}} =  {1\over 3}\sum_{j=1}^{3}{\bf r}_{j} - {\bf r}_{h}~,
{\bf r}_{jl} = {\bf r}_{j} - {\bf r}_{l}, ~{jl}={12},~ {23}, ~{\rm and}~ {31}~.
\label{eq1.1}
\end{equation}
 $\bf R$ has a meaning of the center-of-mass of the exciton coordinate.
 The coordinates ${\bbox \rho}$ and ${\bf r}_{jl}$ are the internal
 variables which are not affected by the translational motion of an exciton.
 Complex coordinates $z_{jl}=x_{jl}+iy_{jl}$, as well as ${\bf r}_{jl}$,
 are not independent. Indeed,
\begin{equation}
{\bf r}_{12} + {\bf r}_{23} + {\bf r}_{31} = 0~.    
\label{eq1.2}                          
\end{equation}
 Despite the fact that the constraint (\ref{eq1.2}) results in some
 complications, the
 introduction of the variables ${\bf r}_{jl}$ enables one to develop the
 theory in a form symmetric in all anyons and, therefore, finally simplifies
 the equations.

Anyons and hole live in two different parallel planes separated by the
 distance $h$. Nevertheless, only the 2D coordinates of Eq.~\ref{eq1.1}
 enter into the exciton wave function. The separation $h$ enters only
 into the Hamiltonian of the anyon-hole interaction derived in
 the Sec.~\ref{subsec:anyhole}.

 The most general form of the pseudo-wave-function of an anyon exciton
 meeting all general requirements is as follows:
\begin{eqnarray}
  \Psi_{L,{\bf k}}({\bf R}, {\bbox \rho}, \{ {\bar{z}_{jl}}\})
  = &&
 \exp\{i{\bf k}{\bf R} + {i\over 2}{\hat z}\cdot ({\bbox \rho}\times {\bf R})
  -{1\over 4}(\bbox{\rho} - {\bf d})^{2}\} 
P_{L}(\dots \bar{z}_{jl} \dots)
 \nonumber \\ && \times
 \prod_{jl}~ (\bar{z}_{jl})^{\alpha}
\exp\{-|{z_{jl}}|^{2}/36\}/{\sqrt{2\pi A}}~,
\label{eq1.3}
\end{eqnarray}
 where the pair of indices, $jl$, takes the values specified in
 Eq.~(\ref{eq1.1}), $A$
 is the normalization area, and $P_L$ is a homogeneous polynomial
 in  coordinates ${\bar z}_{jl}$ of the degree $L$.

 The basic properties of the functions 
 $\Psi_{L,{\bf k}}$ can be checked by inspection. 
 (i) Since the exciton is a neutral entity, it possesses an in-plane momentum
 $\bf k$,\cite{GD} and $\Psi_{L,{\bf k}}$ satisfies the equation of magnetic
 translations:
\begin{equation}
 T_{\bf a}
 \Psi_{L,{\bf k}}({\bf R}, {\bbox \rho}, \{ {\bar{z}_{jl}}\}) =
 \exp [ i ~{\bf a}\cdot {\bf A}({\bbox \rho})]
  \Psi_{L,{\bf k}}({\bf R} - {\bf a}, {\bbox \rho}, \{ {\bar{z}_{jl}}\})
 = {\rm e}^{- i{\bf k}\cdot {\bf a}}
 \Psi_{L,{\bf k}}({\bf R}, {\bbox \rho}, \{ {\bar{z}_{jl}}\})~.
 \label{eq1.4}
\end{equation}
 The parameter ${\bf d} = {\hat z}\times {\bf k}$ is related
 to the dipole moment of the exciton, $(-{\bf d})$. (ii) The function
 $\Psi_{L,{\bf k}}$ belongs to the lowest Landau level. Indeed,
 the nonanalytic factor of it can be shown to have the form
 $\exp\{ - {1\over {12}} \sum_{j} \vert z_{j}\vert^{2} -
 {1\over 4} \vert z_{h}\vert^{2}\}$, whereas the other factors are
 analytic functions of $\bar z_j$ and $z_h$. (iii) The function
 $P_{L}(\dots \bar{z}_{jl} \dots)$ is a homogeneous polynomial of
 the degree $L$ which is symmetric in all coordinates ${\bar {z}}_j$.
 These polynomials form a boson basis,
 and the effect of the fractional statistics is taken into account
 by the factor $ \prod_{jl} (\bar{z}_{jl})^{\alpha}$.\cite{Halp84,wu84}
 (iv) A system of four 
 charged particles in a magnetic field possesses four quantum numbers.
 Two of them are absorbed in the 2D momentum $\bf k$. Two others
 determine the form of the polynomial $P_L$ and are
 internal quantum numbers of an anyon exciton. The operator ${\hat L}_z$
 of the $z$ projection of the angular momentum commutes with the
 Hamiltonian and the square of the momentum, ${\hat {\bf k}}^2$,
 but it does not commute with the projections of ${\hat {\bf k}}$,
 i.e., with ${\hat k}_x$ and ${\hat k}_y$. Therefore, the function 
 $\Psi_{L,{\bf k}}$ chosen in the ${\hat k}_{x}, {\hat k}_y$ representation
 is simultaneously an eigenfunction of ${\hat L}_z$ only for ${\bf k} = 0$,
 and in this limit $L_{z} = -L$.

 Therefore, the quantum numbers of an anyon exciton include the 2D momentum
 $\bf k$ and the projection of the angular momentum,
 $L_{z} = -L$, of the exciton with
 ${\bf k} = 0$. The angular momentum $L$ numerates branches of the 
 exciton spectrum. The fourth quantum number, which will be specified in
 what follows, numerates branches with coinciding values of $L$. The multiple
 branch structure of the anyon exciton spectrum is a direct consequence
 of the charge fractionalization which results in the appearance of the
 internal degrees of freedom of an anyon exciton and of the related
 internal quantum numbers.

By definition, the polynomial $P_L$ is symmetric in  coordinates
 ${\bar z}_j$. To establish the symmetry of it in the symmetric coordinates
 ${\bar z}_{jl}$, one can start with a monomial 
 ${\bar z}_{12}^{l_{3}} {\bar z}_{23}^{l_{1}} {\bar z}_{31}^{l_{2}}$,
 apply to it all operations of the permutation group, and take the sum
 over the group. This transformation
 results in the polynomial
\[
 ({\bar z}_{12}^{l_{3}} {\bar z}_{23}^{l_{1}} {\bar z}_{31}^{l_{2}} +
  {\bar z}_{23}^{l_{3}} {\bar z}_{31}^{l_{1}} {\bar z}_{12}^{l_{2}} +
  {\bar z}_{31}^{l_{3}} {\bar z}_{12}^{l_{1}} {\bar z}_{23}^{l_{2}}) +
 (-)^{l_{1} + l_{2} + l_{3}}
 ({\bar z}_{12}^{l_{3}} {\bar z}_{31}^{l_{1}} {\bar z}_{23}^{l_{2}} +
  {\bar z}_{23}^{l_{3}} {\bar z}_{12}^{l_{1}} {\bar z}_{31}^{l_{2}} +
  {\bar z}_{31}^{l_{3}} {\bar z}_{23}^{l_{1}} {\bar z}_{12}^{l_{2}} ),
\]
which has different properties depending on the parity of 
 $L = l_{1} + l_{2} + l_{3}$. If $L$ is even, the polynomial is a permanent,
 and, therefore, is symmetric in the  coordinates ${\bar z}_{jl}$. 
 However, when $L$ is odd, the polynomial is a determinant which is 
 obviously antisymmetric in the coordinates ${\bar z}_{jl}$ 
 and is non-equal to zero
 only for $l_{1} \neq l_{2} \neq  l_{3}$. For example, for the lowest
 possible value of $L$, $L = 3$, this determinant turns into a
 Vandermonde determinant
\begin{eqnarray}
 W({\bar z}_{12}, {\bar z}_{23}, {\bar z}_{31}) =
 \: \left|
\begin{array}{ccc}
1 & 1 & 1\\
 {\bar z}_{12} & {\bar z}_{23} & {\bar z}_{31}\\
 {\bar z}_{12}^{2} & {\bar z}_{23}^{2} & {\bar z}_{31}^{2}
 \end{array}  \;\right|\; =
 ({\bar z}_{12} - {\bar z}_{23}) ({\bar z}_{23} - {\bar z}_{31})
 ({\bar z}_{31} - {\bar z}_{12})~.
\label{eq1.5}
\end{eqnarray}
 Therefore, $L$-even and $L$-odd polynomials $P_L$ have rather
 different properties. All of them are symmetric in bosonic permutations
 ${\bar z}_{1} \leftrightarrow {\bar z}_{2}$, etc., but they have opposite 
 symmetry with respect to the permutations of the
 ${\bar z}_{12} \leftrightarrow {\bar z}_{23}$ type. 
 To find the explicit form of the polynomials $P_L$, it is convenient
 to introduce  new real coordinates
\begin{equation}
 {\bbox \xi}_{j} = {\bf r}_{j} - {\bf r}_{0}~,~~~ {\bf r}_{0} 
 = {1\over 3}\sum_{l=1}^{3} {\bf r}_{jl}~,
\label{eq1.6}
\end{equation}
 where ${\bf r}_{0}$ is the center-of-mass of the anyon sybsystem. The 
 corresponding complex coordinates are $\zeta_{j} = {\xi}_{jx} + i{\xi}_{jy}$.
 These coordinates are subject of the constraint:
\begin{equation}
 \sum_{j=1}^{3} {\bbox \xi}_{j} =0~, ~~~ \sum_{j=1}^{3} {\zeta}_{j} = 0~.
\label{eq1.7}
\end{equation}

 In these  coordinates the bosonic symmetry of $P_L$ has the usual form,
 and we can apply the fundamental theorem of the theory of symmetric
 polynomials.\cite{BML} According to it,
 $P_{L}({\bar \zeta}_{1}, {\bar \zeta}_{2}, {\bar \zeta}_{3})$
 can be expressed in the unique way as a polynomial in the three
 elementary symmetric polynomials:
\begin{equation}
 p_{1} = {\bar \zeta}_{1} + {\bar \zeta}_{2} + {\bar \zeta}_{3}~,~~
 p_{2} = {\bar \zeta}_{1}{\bar \zeta}_{2} +
 {\bar \zeta}_{2}{\bar \zeta}_{3} + {\bar \zeta}_{3}{\bar \zeta}_{1}~, ~~
 p_{3} = {\bar \zeta}_{1}{\bar \zeta}_{2}{\bar \zeta}_{3}~.
\label{eq1.8}
\end{equation}
The first polynomial is equal to zero, $p_{1} = 0$, because of the constraint
 of Eq.~(\ref{eq1.7}). Therefore, the polynomials $P_L$ are polynomials
 only in $p_{2}$ and $ p_{3}$. The first $L$-even polynomials are:
 $P_0$ = const, $P_{2} \propto p_{2}$, $P_{4} \propto p_{2}^{2}$,
 i.e., there exists only a single  elementary polynomial of
 a given degree $L$. However,
 two elementary polynomials, $p_{2}^{3}$ and $p_{3}^2$, contribute
 to $P_6$. It is easy to check that the number of basis functions
 increases by one each time when $L$ takes values $L = 6m$, where $m$
 is an integer. Therefore, the number of $L$-even polynomials is equal
 to $[L/6]+1$, where  $[L/6]$ is the integral part of $L/6$. All $L$-odd
 polynomials can be obtained by multiplying $L$-even polynomials by $p_3$.
 The latter equals $p_{3} = - {1\over 27} W$ because of Eq.~(\ref{eq1.6}). 

 We are now in position to choose a full basis of polynomials $P_L$
 in symmetric coordinates ${\bar z}_{jl}$.
 These coordinates are most convenient
 for all the following calculations. $L$-even polynomials can be chosen as
\begin{equation}
P_{L,M} = \bar{z}_{12}^{L-4M}\bar{z}_{23}^{2M}\bar{z}_{31}^{2M} +
\bar{z}_{23}^{L-4M}\bar{z}_{31}^{2M}\bar{z}_{12}^{2M} +
\bar{z}_{31}^{L-4M}\bar{z}_{12}^{2M}\bar{z}_{23}^{2M} ~,
\label{eq1.9}
\end{equation}
 where $M = 0,~1,\dots [L/6]$. Polynomials $P_{L,M}$
  are linearly independent, and the
 total number of polynomials with a given $L$ is equal to $[L/6] + 1$.
 All linearly
 independent $L$-odd polynomials can be obtained as:
\begin{equation}
P_{L,M} = W P_{L-3,M}~,~~P_{3,0} = W~.
\label{eq1.10}
\end{equation}
 The total number of them equals $[(L-3)/6] + 1$.

 This choice of polynomials
 determines the full set of quantum numbers in the wave function of
 Eq.~(\ref{eq1.3}) as  $L$, $M$, and $\bf k$. 
 To our best knowledge,
 in the previous studies only the $L$-even polynomials have been
 taken into account. \cite{ET92} When choosing polynomials $P_{L,M}$,
 we have not imposed the hard core constraint and
 defer the discussion of the related properties to what follows.

It is an important feature of the AEM that the wave functions (and, therefore,
electron densities, etc.) of all eigenstates with $L\leq 5$ and also $L=7$
 are completely 
 determined by the symmetry requirements. They do not depend on the
 specific form of the Hamiltonian and, in particular, on $h$.

\section{ The Schroedinger equation}
\label{sec:Hamilt} 

 In this section we calculate the Hamiltonian of the AEM
 in the $\Psi_{L, M, {\bf k}}$
 basis as the matrix of a point charge Coulomb interaction.  
 It means, that we neglect form-factors of quasielectrons which
 have a scale of several magnetic lengths and are known only
 approximately.\cite{CP,Morf,GH94} We postpone the discussion of inaccuracy
 originating from this approximation to Sec.~\ref{sec:comparison}.

 Unfortunately, functions $\Psi_{L, M, {\bf k}}$ are orthogonal only
 in  quantum numbers $L$ and $\bf k$. As a result, the scalar products
  $\langle \Psi_{L, M, {\bf k}}, \Psi_{L, M', {\bf k}}\rangle \neq 0$
 for $M \neq M'$,
 and the matrix ${\hat B}$ of these scalar products is block-diagonal.
 The size of blocks is equal to 1 for $L < 6$ and $L=7$ and increases by 1
 each time when $L$ increases by 6. With a non-diagonal matrix  ${\hat B}$
 the Schroedinger equation has a form:
\begin{equation}
 {\hat H} \chi = \varepsilon {\hat B} \chi~,
\label{eq2.1}
\end{equation}
 and one has to find matrices ${\hat H}$ and ${\hat B}$. To perform the
 calculations, it is convenient to employ variables $\bf R$, $\bbox \rho$,
 and three ${\bf r}_{ij}$ and to take into account the constraint
 of Eq.~(\ref{eq1.2}) by the usual transformation:
\begin{equation}
 \delta({\bf r}_{12} + {\bf r}_{23} + {\bf r}_{31}) =
 \int {d{\bf f}\over {(2\pi)^{2}}} 
 \exp \{ i {\bf f}\cdot ({\bf r}_{12} + {\bf r}_{23} + {\bf r}_{31})\}~.
\label{eq2.2}
\end{equation}
 It adds the new variable $\bf f$, but all calculations become symmetric
 in anyon variables. The Jacobian of the transformation is equal to 1.

 The Hamiltonian is diagonal in $\bf k$, therefore, we write out only
 the diagonal in $\bf k$ matrix elements. For ${\bf k} = 0$, the Hamiltonian
 is also diagonal in $L$, and  Eq.~(\ref{eq2.1}) acquires a
 block-diagonal form.

 Since all terms in the polynomials $P_{L, M, {\bf k}}$~,
 Eqs.~(\ref{eq1.9}) and (\ref{eq1.10}),  have the same form,
 we concentrate in what follows on the matrix elements taken in the basis
 of the functions $\Psi_{\{n\}{\bf k}}$~:
\begin{eqnarray}
  \Psi&&_{\{n\}{\bf k}}({\bf R}, {\bbox \rho}, \{{\bar z}_{jl}\}) \nonumber \\
  =&&
  \exp\{i{\bf k}{\bf R} + {i\over 2}{\hat z}\cdot ({\bbox \rho}\times {\bf R})
  -{1\over 4}(\bbox{\rho} - {\bf d})^{2}\}
 {\bar z}_{12}^{n_{3}+\alpha}{\bar z}_{23}^{n_{1}+\alpha}
 {\bar z}_{31}^{n_{2}+\alpha}
 \exp\{-\sum_{jl}|{z_{jl}}|^{2}/36\}/{\sqrt{2\pi A}}~.
\label{eq2.3}
\end{eqnarray}
 Here polynomials $P_{L, M}$ are substituted by monomials, and $\{n\}$
 denotes a set of quantum numbers $n_{1}, n_{2}$ and $n_3$.
 In the next parts of this Section we describe in some detail the technique
 for performing different types of integrals. 

\subsection{Nonorthogonality matrix elements}
\label{subsec:Nonorth}

The scalar products of functions $\Psi_{\{n\}{\bf k}}$, when written
 in the variables $\bf R$, $\bbox \rho$, and ${\bf r}_{jl}$, have the form:
\begin{eqnarray}
 B_{ \{n\}  \{n'\}} =
 \langle &&\{n\} \mid \{n'\} \rangle  \nonumber \\= &&
 \int d{\bf R} \int d{\bbox \rho} \int {d{\bf f}\over {(2\pi)^{2}}}
 \int d{\bf r}_{12} d{\bf r}_{23} d{\bf r}_{31}
 {\bar \Psi}_{\{n\}{\bf k}} \Psi_{\{n'\}{\bf k}}
 \exp\{ i{\bf f}\cdot ({\bf r}_{12} + {\bf r}_{23} + {\bf r}_{32})\}~.
\label{eq2.4}
\end{eqnarray}
 Integration over $\bf R$ and the Gaussian integration over $\bbox \rho$
are straightforward. Since $\Psi_{\{n\}{\bf k}}$ is multiplicative
 in the variables ${\bf r}_{jl}$, one can rewrite Eq.~(\ref{eq2.4})
 in the form:
\begin{equation}
 \langle \{n\}  \mid \{n'\} \rangle =
 \int {d{\bf f}\over {(2\pi)^{2}}}
 \prod_{j=1}^{3} M_{n_{j}n'_{j}}^{(\alpha)}({\bf f})~,
\label{eq2.5}
\end{equation}
where
\begin{equation}
 M_{mm'}^{(\alpha)}({\bf f}) =
 \int d{\bf r} ~r^{2\alpha}~ z^{m} {\bar z}^{m'}~
 {\rm e}^{-r^{2}/18 + i{\bf f}\cdot {\bf r}}~.
\label{eq2.6}
\end{equation}
 Since 
\begin{equation}
 \int_{0}^{2\pi} d{\varphi}~ {\rm e}^{\pm im\varphi + ifr\cos{\varphi}}
 = 2\pi i^{\mid m \mid} J_{\mid m \mid} (fr)~,
\label{eq2.7}
\end{equation}
 where $ J_{\mid m \mid}(fr)$ is a Bessel function,
 the angle integration 
  in (\ref{eq2.6}) results in:
\begin{equation}
 M_{mm'}^{(\alpha)}({\bf f}) = 2\pi i^{\mid m-m' \mid}
 \exp [i(m-m'){\varphi}_{\bf f}]~
 {\cal M}_{mm'}^{(\alpha)}(t)~,
\label{eq2.8}
\end{equation}
 where $t=9f^{2}/2$,~ ${\varphi}_{\bf f}$ is the azimuth of $\bf f$,  and
 \begin{equation}
 {\cal M}_{mm'}^{(\alpha)}(t)
 = \int_{0}^{\infty} dr ~r^{1 + 2\alpha + m + m'} {\rm e}^{- r^{2}/18}
 J_{\mid m-m' \mid}(fr)~.
\label{eq2.9}
\end{equation}
This integral can be expressed in terms of the confluent hypergeometric
 function $\Phi(\beta, \gamma; t)$ as\cite{GRyz}
 \begin{eqnarray}
 {\cal M}_{mm'}^{(\alpha)}(t)&& ~= {\Gamma}( \max\{m, m'\} + \alpha + 1)~
 { 2^{{{\mid m+m'\mid}/2} + \alpha}~ 3^{ {\mid m+m'\mid} + 2(\alpha + 1)}
\over {{\mid m-m'\mid}!} }~ t^{{\mid m-m'\mid}/2} \nonumber \\ &&  \times~
 \Phi(\max \{m, m'\} + \alpha + 1, {\mid m-m'\mid} + 1; -t)~.
\label{eq2.10}
\end{eqnarray}
 Here $\max \{m, m'\}$ is the larger of the integers $m$ and $m'$.
 After the integration over ${\varphi}_{\bf f}$ in Eq.~(\ref{eq2.5}),
 the coefficients
 $\langle \{n\}  \mid \{n'\} \rangle$
 take the form:
 \begin{equation}
 \langle \{n\}  \mid \{n'\} \rangle =
 {\delta}_{nn'} (2\pi/3)^{2} \int_{0}^{\infty} dt
 \prod_{j=1}^{3} i^{\mid n_{j}- n'_{j}\mid}
 {\cal M}^{(\alpha)}_{n_{j} n'_{j}} (t)~,
\label{eq2.11}
\end{equation}
 where
 \begin{equation}
 n = n_{1} + n_{2} + n_{3}~.
\label{eq2.12}
\end{equation}
 Therefore, the scalar product
 $\langle \{n\}  \mid \{n'\} \rangle$
 of two functions
 $\Psi_{\{n\}{\bf k}}$ is reduced to an one-fold integral from the
 product of three confluent hypergeometric functions.
 Matrix elements $\langle \{n\}  \mid \{n'\} \rangle$
 do not depend on $\bf k$.
  Scalar products
 $\langle L, M  \mid L^{'}, M^{'} \rangle$
of two functions $\Psi_{L, M, {\bf k}}$~, which include polynomials $P_{L,M}$,
 are linear combinations of the coefficients
  $\langle \{n\}  \mid \{n'\} \rangle$. 
 They do not depend on $\bf k$, too.
 The coefficients of these combinations can be found
 from the equations (\ref{eq1.5}),
 (\ref{eq1.9}), and (\ref{eq1.10}),
 but the final expressions are rather cumbersome,
 especially for $L$-odd polynomials. Therefore, we do not write out here
 their explicit form.

 Integrals (\ref{eq2.11}) can be simplified for $\alpha = 0$. Indeed,
 in this case the Kummer transformation\cite{GRyz}
 \begin{equation}
 \Phi(\beta,    \gamma; t) = {\rm e}^{t} \Phi(\gamma - \beta, \gamma; -t)
\label{eq2.13}
\end{equation}
 results in a $\Phi$ function with the first
 parameter $(\gamma - \beta)$ equal to a negative integer. This
 function reduces to a polynomial, and $\Phi(\beta,    \gamma; t)$ to a
 polynomial multiplied by ${\rm e}^{t}$.  Therefore, the integral in
 Eq.~(\ref{eq2.11}) for $\langle \{n\}  \mid \{n'\} \rangle$
  can be performed exactly. This transformation
 highly simplifies calculations. Indeed, in the large $h$ region,
 where the criterion
 of the applicability of the AEM is satisfied,
 the statistical parameter $\alpha$ can be neglected as it is shown
 in Sec.~\ref{sec:Energy} below.

\subsection{Anyon-anyon interaction}
\label{subsec:anyany}

 The Hamiltonian of the anyon-anyon interaction is:
 \begin{equation}
 {\hat V}_{aa} = {1\over 9} \{ {\mid {\bar z}_{12} \mid}^{-1} +
 {\mid {\bar z}_{23} \mid}^{-1} + {\mid {\bar z}_{31} \mid}^{-1} \}~.
\label{eq3.1}
\end{equation}
 Matrix elements of ${\hat V}_{aa}$ in the basis of the functions
 $\Psi_{\{n\}{\bf k}}$ can be calculated by analogy with
 the matrix elements
$\langle \{n\}  \mid \{n'\} \rangle$.
 The denominators
 ${|{\bar z}_{jl}|}^{-1}$ lower the power of $r$ by 1 in
one of the ${\cal M}_{n_{j}n'_{j}}^{(\alpha)}({\bf f})$ factors
 entering in Eq.~(\ref{eq2.5}). The final expression
 for the matrix element is:
 \begin{eqnarray}
 \langle  \{n\}&& ~\mid {\hat V}_{aa} \mid \{n'\} \rangle
 = 
  {1\over 9} \int {d{\bf f}\over {(2\pi)^{2}}}
 \{M_{n_{1}n'_{1}}^{(\alpha  - {1\over 2})}({\bf f})
 M_{n_{2}n'_{2}}^{(\alpha )}({\bf f}) 
 M_{n_{3}n'_{3}}^{(\alpha)}({\bf f}) \nonumber \\&& +
 M_{n_{1}n'_{1}}^{(\alpha)}({\bf f})
 M_{n_{2}n'_{2}}^{(\alpha - {1\over 2})}({\bf f})
 M_{n_{3}n'_{3}}^{(\alpha)}({\bf f}) +
 M_{n_{1}n'_{1}}^{(\alpha)}({\bf f})
 M_{n_{2}n'_{2}}^{(\alpha)}({\bf f})
 M_{n_{3}n'_{3}}^{(\alpha - {1\over 2})}({\bf f}) \}~.
\label{eq3.2}
\end{eqnarray}
 Like the nonorthogonality matrix elements,
 matrix elements of ${\hat V}_{aa}$ also do not depend on the momentum $\bf k$.
 For this reason, the matrix of the operator ${\hat V}_{aa}$ is diagonal
 in the angular momentum, $L$, for arbitrary values of the momentum $\bf k$.

\subsection{Anyon-hole interaction}
\label{subsec:anyhole}

 The Hamiltonian of the anyon-hole interaction has a form:
\begin{equation}
 {\hat V}_{ah} = - ~{1\over 3}~\sum_{j=1}^{3} {\hat V}_{jh}~, ~~
 {\hat V}_{jh}({\bf r}_{jh}) = r_{jh}^{-1}~,
\label{eq4.1}
\end{equation}
 where the three-dimensional anyone-hole separation
 ${\bf r}_{1h}$ should be expressed
 in terms of the difference coordinates:
\begin{equation}
 {\bf r}_{1h} = {\bbox \rho} + {1\over 3} ({\bf r}_{12} - {\bf r}_{31}) +
 {\hat z} h ~.
\label{eq4.2}
\end{equation}
 Similar equations hold for ${\bf r}_{2h}$ and ${\bf r}_{3h}$.

 It is convenient to introduce the Fourier image
 \begin{equation}
 {\hat V}_{1h}({\bf r}_{1h}) =
 - ~{1\over 3} \int { d{\bf q}\over {(2\pi)^{2}}}
 V_{ah}(q) \exp \{ i{\bf q}\cdot {\bbox \rho} + {i\over 3} {\bf q}\cdot
 ({\bf r}_{12} - {\bf r}_{31}) \}~,
\label{eq4.3}
\end{equation}
 where $V_{ah}(q) = ({2\pi}/q)\exp(-qh)$.
 The integrations over $\bf R$ and $\bbox \rho$ in the matrix elements
 of ${\hat V}_{1h}({\bf r}_{1h})$ can be performed in the same way as
 in Eq.~(\ref{eq2.4}), and the result can be expressed in terms of the
 coefficients $M_{mm'}^{(\alpha)}$~, Eq.~(\ref{eq2.6}), as:
\begin{eqnarray}
 \langle \{n\} \mid {\hat V}_{1h}\mid \{n'\} \rangle =&&
 -~{1\over 3}\int { d{\bf f}\over {(2\pi)^{2}}}
 \int { d{\bf q}\over {(2\pi)^{2}}}~
 V_{ah}(q) ~{\rm e}^{ - q^{2}/2 + i{\bf k}\cdot{\bf q} } \nonumber \\
 && \times ~M_{n_{1}n'_{1}}^{(\alpha)}({\bf f})
 ~M_{n_{2}n'_{2}}^{(\alpha)}({\bf f}-{\bf q}/3)
 ~M_{n_{3}n'_{3}}^{(\alpha)}({\bf f}+{\bf q}/3)~.
\label{eq4.4}
\end{eqnarray}
 After the angular integration, the factors $M_{n_{j}n_{j}^{'}}^{(\alpha)}$
 take the form of Eq.~(\ref{eq2.8}), and Eq.~(\ref{eq4.4}) can be 
 rewritten as:
\begin{eqnarray}
  \langle &&\{n\} \mid {\hat V}_{1h}\mid \{n'\} \rangle =
 - { {(2\pi)^{2}}\over 3} \prod_{j=1}^{3} i^{\mid n_{j}-n'_{j}\mid}
 \int { d{\bf f}\over {(2\pi)^{2}}}
 \int { d{\bf q}\over {(2\pi)^{2}}}~
 V_{ah}(q) ~{\rm e}^{ - q^{2}/2 + i{\bf k}\cdot{\bf q} } \nonumber \\
 &&\exp \{ i{\varphi}_{\bf f}(n_{1}-n'_{1}) + 
    i{\varphi}_{-}(n_{2}-n'_{2}) + i{\varphi}_{+}(n_{3}-n'_{3})\} \nonumber \\
 && {\cal M}_{n_{1}n'_{1}}^{(\alpha)}(9f^{2}/2)~
 {\cal M}_{n_{2}n'_{2}}^{(\alpha)}(9f_{+}^{2}/2)~	
 {\cal M}_{n_{3}n'_{3}}^{(\alpha)}(9f_{-}^{2}/2)~.
\label{eq4.5}
\end{eqnarray}
  Vectors ${\bf f}_{\pm}$ are defined by the equation
 \begin{equation}
 {\bf f}_{\pm} = {\bf f} \pm {\bf q}/3~.
\label{eq4.6}
\end{equation}
  One can perform one angular integration in Eq.~(\ref{eq4.5})
 if the phase ${\varphi}_{\bf f}$ is eliminated by change in the variables:
 \begin{equation}
 \varphi = {\varphi}_{\bf f} - {\varphi}_{\bf q}~,~~
 \psi_{\pm} = {\varphi}_{\pm} - {\varphi}_{\bf f}~.
\label{eq4.7}
\end{equation}
 Here $\varphi_{\pm}$ are phases of vectors ${\bf f}_{\pm}$.

 After the integration over ${\varphi}_{\bf q}$, 
  Eq.~(\ref{eq4.4}) takes its final form:
\begin{eqnarray}
 \langle \{n\} \mid {\hat V}_{1h}\mid \{n'\} \rangle = &&
 - {1\over 3}~ i^{\mid n - n' \mid} 
 \prod_{j=1}^{3} i^{\mid n_{j}-n'_{j}\mid}
  \int_{0}^{\infty} df f \int_{0}^{\infty} dq~q 
  ~V_{ah}(q) ~{\rm e}^{ - q^{2}/2 } J_{\mid n - n' \mid}(kq) \nonumber \\
 &&\times~\int_{0}^{2\pi} d{\varphi} ~\exp\{ i{\varphi}(n-n') +
 i\psi_{-}(n_{2}-n'_{2}) + i\psi_{+}(n_{3}-n'_{3}) \} \nonumber \\
 &&\times~ {\cal M}_{n_{1}n'_{1}}^{(\alpha)}(9f^{2}/2)~
 {\cal M}_{n_{2}n'_{2}}^{(\alpha)}(9f_{+}^{2}/2)~        
 {\cal M}_{n_{3}n'_{3}}^{(\alpha)}(9f_{-}^{2}/2)~.
\label{eq4.8}
\end{eqnarray}
 Here the direction of $\bf k$ was chosen along the $x$ axis to
 eliminate the complex phase from the matrix element (\ref{eq4.8}).
 Explicit expressions
 of $\psi_{\pm}$ and $f_{\pm}$ in terms of the integration variables are
 as follows:
 \begin{equation}
 {\rm e}^{i\psi_{\pm}} = (f \pm {q\over 3}{\rm e}^{-i{\varphi}})/f_{\pm}~,
 ~~ f_{\pm}^{2} = f^{2} + (q/3)^{2} \pm {2\over 3} f q \cos {\varphi} ~.
\label{eq4.9}
\end{equation}
One can check by inspection that Eq.~(\ref{eq4.8}) is symmetric in
 the indices $n_{2}, n'_{2}$ and $ n_{3}, n'_{3}$, and that matrix
 elements are real for all sets of the quantum numbers $\{n_{j}\}, 
 \{n'_{j}\}$ compatible with our choice of the polynomials $P_{L, M}$,
 Eqs.~(\ref{eq1.9}) and (\ref{eq1.10}). Matrix elements
 of the operators ${\hat V}_{2h}$ and ${\hat V}_{3h}$ can be written
 by analogy with Eq.~(\ref{eq4.8}).

 Equations (\ref{eq3.2}) and (\ref{eq4.8}) for the Hamiltonian $\hat H$
 and Eqs.~(\ref{eq2.4}) and (\ref{eq2.11})
 for the matrix $\hat B$ determine completely the
 Schroedinger equation (\ref{eq2.1}).

\section{ Energy spectrum}
\label{sec:Energy}

 For ${\bf k} = 0$ the operator of the anyon-hole interaction ${\hat V}_{ah}$
 becomes diagonal in the angular momentum $L$. Therefore, the Schroedinger
  equation (\ref{eq2.1}) is also diagonal in $L$ since ${\hat V}_{aa}$
 and $\hat B$ are diagonal in $L$ for arbitrary ${\bf k}$. It was shown in
 Sec.~\ref{sec:functions} that for each value of $L$ such that $L\leq 5$
 or $L = 7$ there exists a single eigenfunction, and it does not
 depend on $h$. For these values of $L$ the equations of Sec.~\ref{sec:Hamilt}
 immediately give the energies of ${\bf k}=0$ states. 
  When the number of $P_{L, M}$ polynomials with
 a given value of $L$ becomes two or more, equations of Sec.~\ref{sec:Hamilt}
 give the coefficients of secular equations of the second, third, etc.,
 order which determine the $h$ dependent eigenfunctions and eigenvalues.

 The order in which  ${\bf k} = 0$ levels
 are arranged changes with increasing $h$.
 The main regularities can be understood using
 classical arguments. In the classical limit,
 which is achieved for a large exciton size, the exciton ground
 state takes the shape of an equilateral triangle with anyons in the 
 vertices and a hole in the center. 
 The anyon-anyon distance in this triangle, $r_{12}$,
 found from the minimum of the electrostatic energy, is equal to
\begin{equation}
 r_{12} = \sqrt{3/2} h~.
\label{eq5.1}
\end{equation}
 It will be shown below that triangular configurations are  described by
 the polynomials $P_{6M, M}$ and $WP_{6M, M}$. A
 straightforward calculation  based on equations (\ref{eq1.3}), 
 (\ref{eq1.9}) and (\ref{eq1.10}) 
 shows that a mean-square value of the inter-anyon
 distance in these states is equal to
\begin{equation}
 <r_{12}^{2}> = 6(L + 2 + 3\alpha)~.
\label{eq5.2}
\end{equation}
 Comparing equations (\ref{eq5.1}) and (\ref{eq5.2}) one finds that
 \begin{equation}
 r_{12} \sim h \sim L^{1/2}~
\label{eq5.3}
\end{equation}
 for $L\gg 1$. 
 Therefore, with increasing $h$ both the size, $r_{12}$, and the angular
 momentum, $L$, of the exciton ground state increase. This means that
 the order of the energy levels changes, and the bottom state possesses
 the angular momentum $L \sim h^{2}$. Since for ${\bf k} = 0$ the
 Hamiltonian is diagonal in $L$, the level interchange occurs usually
 as a level crossing.
 
\begin{figure}
\begin{center}
\includegraphics{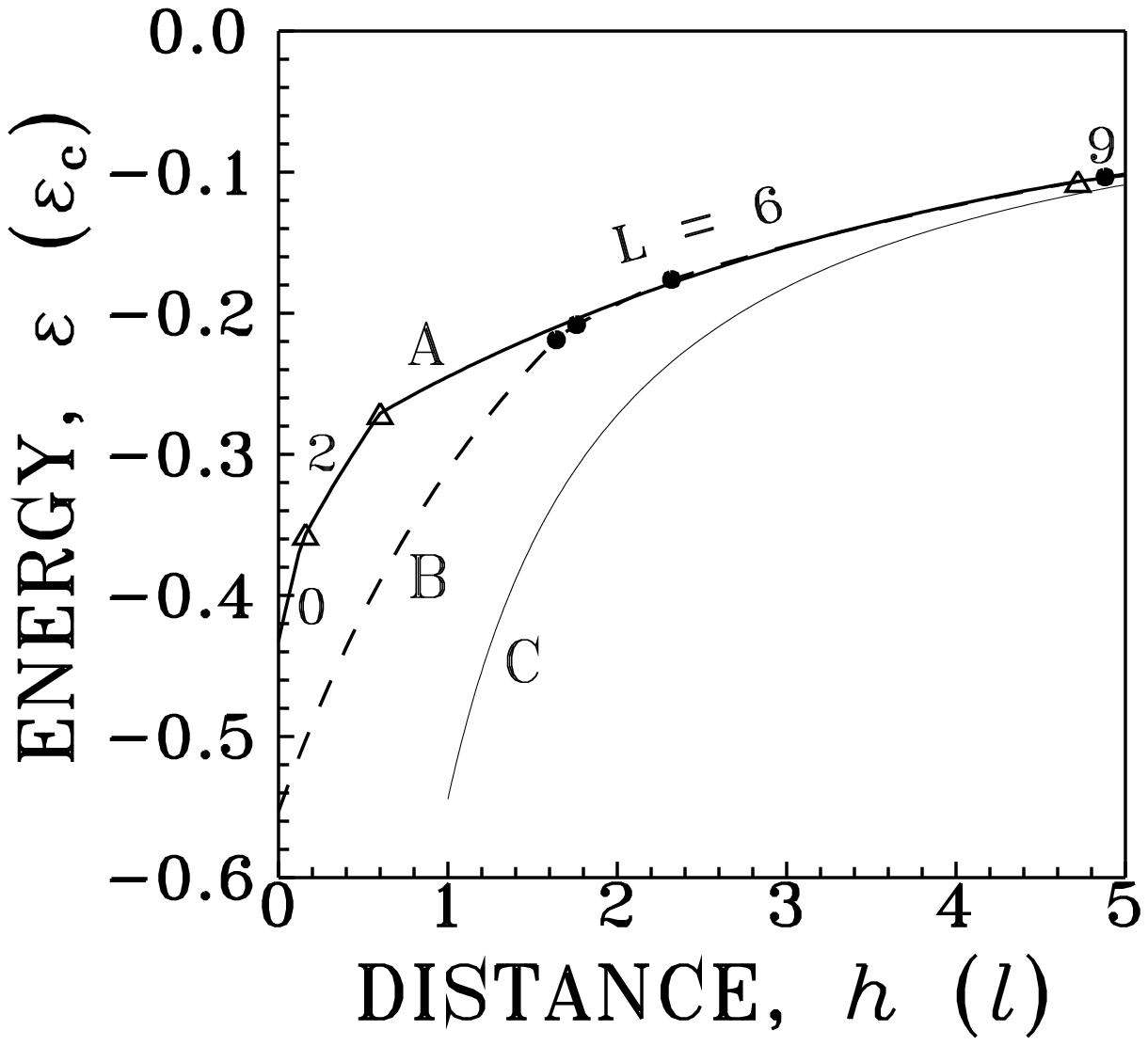}
\end{center}
\vskip 16truecm
 \caption{ Ground state energy, $\varepsilon ({\bf k}=0)$,
 plotted versus separation, $h$, between electron and hole confinement
 planes. A and B - statistical charges $\alpha = -1/3$ and $\alpha = 0$,
 respectively. A - triangles show the points where the angular momentum
 $L$ changes from zero to 2, and then to 6 and 9. For the quantum states
 separated  by full dots on the curve B see Fig.~\ref{fig4}. 
 C - classical limit.}
\label{fig1}
\end{figure}

 In Fig.~\ref{fig1} the ground state energy is plotted as a function of 
 $h$ for two values of the statistical charge, $\alpha = - 1/3$ and
 $\alpha = 0$, by curves A and B, respectively. 
 It is seen that both curves show the same gross features,
 including increase of the ground-state energy and the
 angular momentum with $h$. However, fine details are very different
 in the $h < 2$ region.
 We do not discuss these differences in more detail since they
 are expected to be sensitive to anyon form-factors\cite{CP,Morf,GH94}
 which were not taken into account in our calculations.
 However, since curves A and B  practically coincide for $h\agt 2$,
 we believe that in this region the AEM provides reliable results.
 In what follows we restrict ourselves with this region and
 neglect the statistical charge, i.e., consider the bosonic model, $\alpha = 0$.
 All data below are presented for this model.
 It was shown in Sec.~\ref{subsec:Nonorth}, Eq.~(\ref{eq2.13}),
 that for $\alpha = 0$  matrix elements
 ${\cal M}_{mm'}^{(\alpha)}(t)$ can be expressed in terms of elementary
 functions. This fact permits one to reduce the three-fold integral
 of Eq.~(\ref{eq4.8}) to the one-fold integral:
\begin{equation}
<LM{\bf k}|V_{ah}|L'M'{\bf k}>= - \int_{0}^{\infty}\exp(-3q^{2}/2-qh)~
J_{|L-L'|}(kq)~Q_{LM,L'M'}(q)~ dq~,
\label{eq5.4}
\end{equation}
where 
functions $Q_{LM,L'M'}(q)$, real and symmetric in indices, are
polynomials in $q$. The lower polynomials are of a simple form:
$Q_{00,00}=1,~Q_{20,00}=q^{2}/2,~Q_{20,20}=1-q^{2}+q^{4}/4$. Application
 of the equation (\ref{eq5.4}) highly simplifies all computations.

 For comparison, in Fig.~\ref{fig1} is also shown the energy 
 $\varepsilon_{\rm cl} = - (2/3)^{3/2}/h$ 
 calculated in the classical limit, Eq.~(\ref{eq5.1}).
 It is seen that in the region $h\alt 4$ it differs considerably 
 from the exact quantum data.

\begin{figure}
\begin{center}
\includegraphics{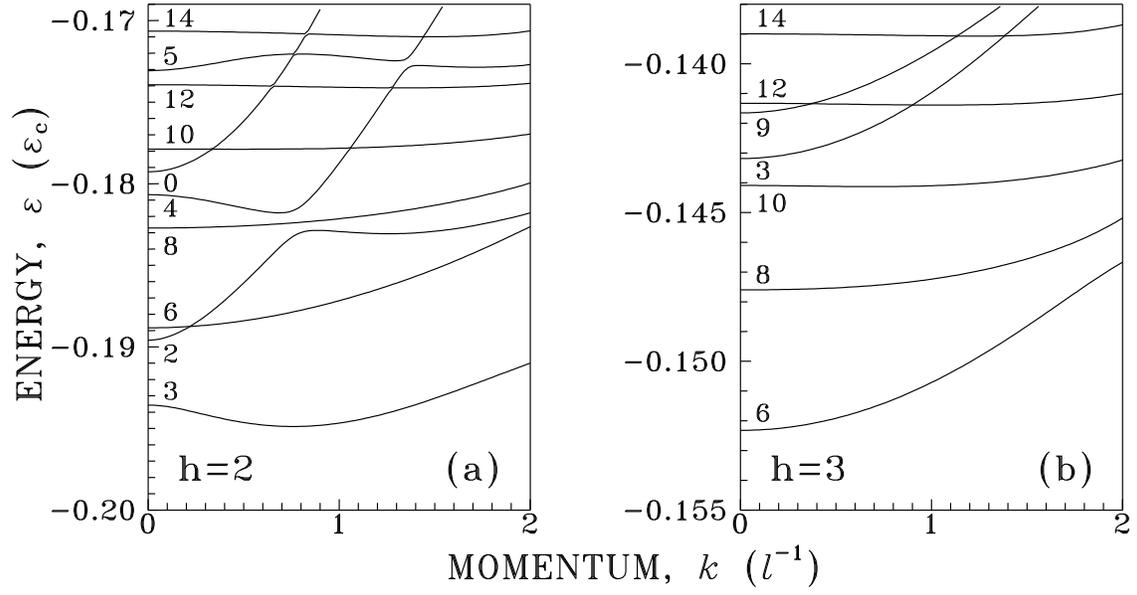}
\end{center}
\vskip 15truecm
 \caption{ Anyon exciton dispersion law $\varepsilon(k)$
 for two values of $h$.  For $h = 2$, the negative dispersion arises
 because of the mutual repulsion of $L = 2$ and $L = 3$ branches.
 Level splitting near avoided crossings becomes tiny with increasing $h$.
 Numbers show $L$ values. $h$ in units of $l$.
 }
\label{fig2}
\end{figure}

 Since for ${\bf k} \neq 0$ the term ${\hat V}_{ah}$ is non-diagonal
 in $L$, the dispersion law $\varepsilon(k)$ can be found only numerically. 
 In Fig.~\ref{fig2} it is shown for two values of $h$. 
 The basis of polynomials used in computations included 22 $L$-even 
 polynomials with $L \leq 18$, Eq.~(\ref{eq1.8}), and 22 $L$-odd  
 polynomials with $L \leq 21$, Eq.~(\ref{eq1.9}). 
 The following regularities are distinctly seen.
 As it was argued above, the levels with higher $L$
 values draw closer to the spectrum bottom  with increasing $h$.
 The level interchange 
 manifests itself as avoided level crossings.
 The level splitting near these crossings
 increases with $k$ and decreases with the difference $|L-L'|$.
 These regularities
 can be understood if one takes into account that the Bessel function
 in the integrand of Eq.~(\ref{eq5.4}) shows the power-law behavior,
 $J_{\mid L-L'\mid}(kq)\propto (kq)^{\mid L-L'\mid}$, in the small $k$
 region. In particular, for ${\mid L-L'\mid} = 1$ the interaction of
 two branches can result in the negative exciton effective mass for
 small $k$ values. Negative exciton dispersion can appear even in
 the ground state as it can be seen in Fig.~\ref{fig2}a. 
 It is interesting to mention that exciton dispersion near 
 ${\bf k} = 0$ is always positive in the two-semion problem.\cite{RP93}

 The above results show that the charge fractionalization 
 determines both i) the basic multiple-branch structure of
 the exciton energy spectrum and ii) numerous specific features of the spectrum
 including the $h$ dependence of the arrangement of the branches,
 avoided branch intersections, etc.

\section{Electron density}
\label{sec:density}

The distribution of the electron density, $D_{\lambda}({\bf r}, {\bf k})$,
 around a hole can be found from the equation:
\begin{equation}
 D_{\lambda}({\bf r}, {\bf k}) =
 \langle ~ {1\over 3} \sum_{j=1}^{3} \delta({\bf r}_{j} - {\bf r}_{h} - {\bf r})
 ~ \rangle _{\lambda {\bf k}}~,
\label{eq6.1}
\end{equation}
 where the averaging is performed over the quantum state
 ($\lambda, {\bf k}$), where $\lambda$ numerates exciton branches. The
 density $D_{\lambda}({\bf r}, {\bf k})$ is exactly the quantity
 which i) permits one to check reliability of the model and  in which
 ii) the specific pattern of the charge fractionalization manifest
 itself in a rather spectacular way. It is one of the basic criteria
 of the AEM that the excess charge density is small 
 compared with the density of the IQL, $\nu/{2\pi}$, for $\nu < 1/2$
 (and compared with $(1-\nu)/{2\pi}$ for $\nu > 1/2$).
 Therefore, for $\nu = 1/3$
 and $\nu = 2/3$ IQL's one can expect that the AEM becomes applicable
 only when
 $D_{\lambda} \ll 1/6\pi \approx 0.05$.
 This criterion will be applied in what follows.

The explicit expression for $D_{\lambda}$ can be obtained in the same
 way as Eq.~(\ref{eq5.4}). Indeed, the operator ${\hat V}_{ah}$
 and the operator of the electron density of Eq.~(\ref{eq6.1})
 depend on the same arguments, ${\bf r}_{j}-{\bf r}_{h}$.
 Therefore, the integrands
 differ only in the substitution $V_{ah}(q)$ by the Fourier image
 of $\delta({\bf r}_{j}-{\bf r}_{h})$. The final expression is:
\begin{eqnarray}
D_{\lambda}({\bf r}, {\bf k}) = && {1\over{2\pi}} \sum_{LM,L'M'}
 \cos[(L-L')\theta] \nonumber \\ &&\times
\int_{0}^{\infty}{ dq~q~\exp(-3q^{2}/2)~J_{|L-L'|}(q|{\bf r}-{\bf d}|)}~
Q_{LM,L'M'}(q)~ {\bar \chi}_{L'M'}^{\lambda}({\bf k})
 \chi_{LM}^{\lambda}({\bf k})~,
\label{eq6.2}
\end{eqnarray}
where $\theta$ is the angle between the vectors
${\bf d} - {\bf r}$ and ${\bf d}$. $D_{\lambda}(0 ,{\bf k})$
 shows the electron density on the hole.

\begin{figure}
\begin{center}
\includegraphics{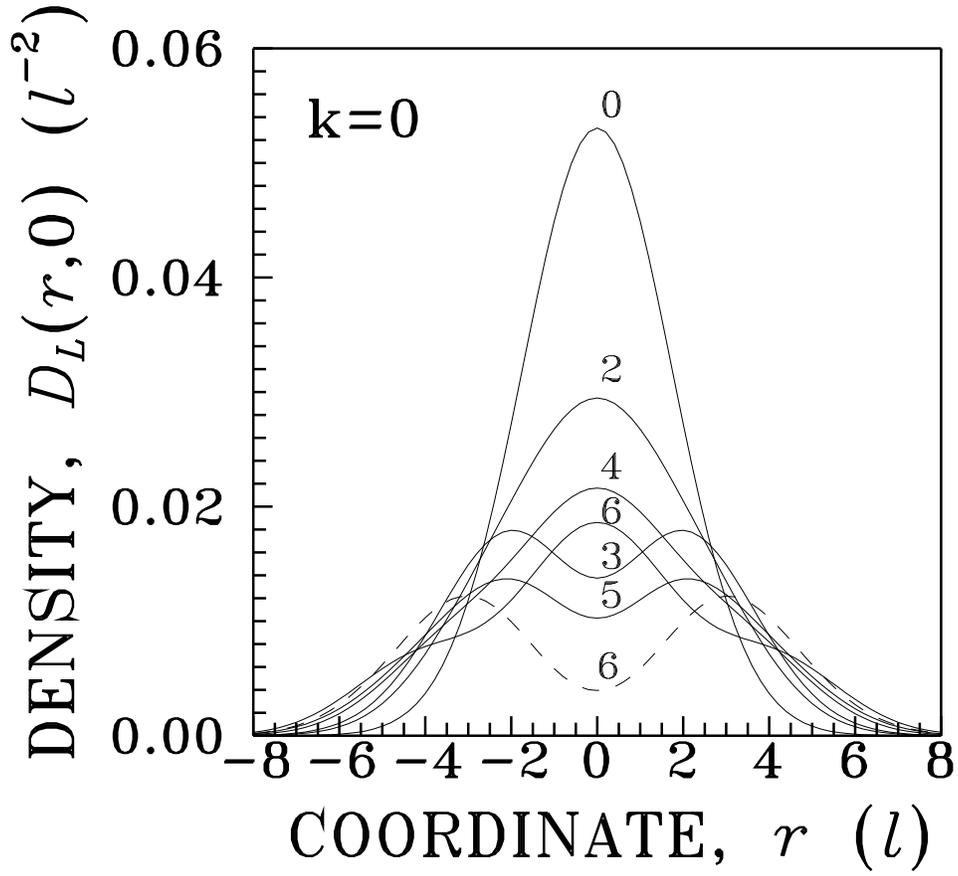}
\end{center}
\vskip 16.0truecm
 \caption{ Axisymmetric electron density distributions $D_{L}(r,0)$
for an ${\bf k} = 0$ exciton
for the states with $L\leq 6$. Two $L=6$ states are shown for
$h=0$; the density distribution in the lower state is shown by a solid line.
 Numbers show $L$ values.
 }
\label{fig3}
\end{figure}

For ${\bf k}=0$, the density distribution $D_{\lambda}(r,0)$ is
 shown in Fig.~\ref{fig3} for $L \leq 6$. In this case $\lambda$ can be
 completely identified by the
 index $L$ for $L\leq 5$, but there are two functions for $L = 6$.
  For $L \leq 5$ the energies, the eigenfunctions, and therefore also
 the densities $D_L$, do not depend on $h$.
 The $L = 0$ state has a high density;
  $D_{0}(0,0) > 0.05$. Therefore, the shape of the curve $D_{0}(r, 0)$
 can not be reliable. Nevertheless, it is remarkable that numerical
calculations performed in the spherical geometry
 for $\nu = 1/3$ \cite{CQ94,APR,ZB95,dens} and $\nu = 2/3$ IQL's\cite{AR2/3}
 convincingly show that for $h \alt 2$ the spectrum bottom is made
 by the $L = 0$ exciton having the electron density which is very
 close to the Fermi limit, $1/2\pi$.\cite{APR,AR-PR}
  Functions $D_{L,M}$ become broader with increasing $L$,
 and $D_{L,M}(0,0)$ decreases.
 The state $L=3$ from which $L$-odd polynomials originate is especially
 remarkable because it is the first to show a crater-like density
 distribution with a minimum at $r=0$. This minimum
 is a signature of the charge fractionalization, since for ordinary
 magnetoexcitons the density shows a maximum at $r = 0$. We will discuss
 the properties of this state in more detail in Sec.~\ref{sec:correlation}.
 For $L=6$ there are two eigenfunctions; they depend on $h$. 
 In Fig.~\ref{fig3} they are shown for $h=0$; 
 the lower energy component is drawn by a solid line.

\begin{figure}
\begin{center}
\includegraphics{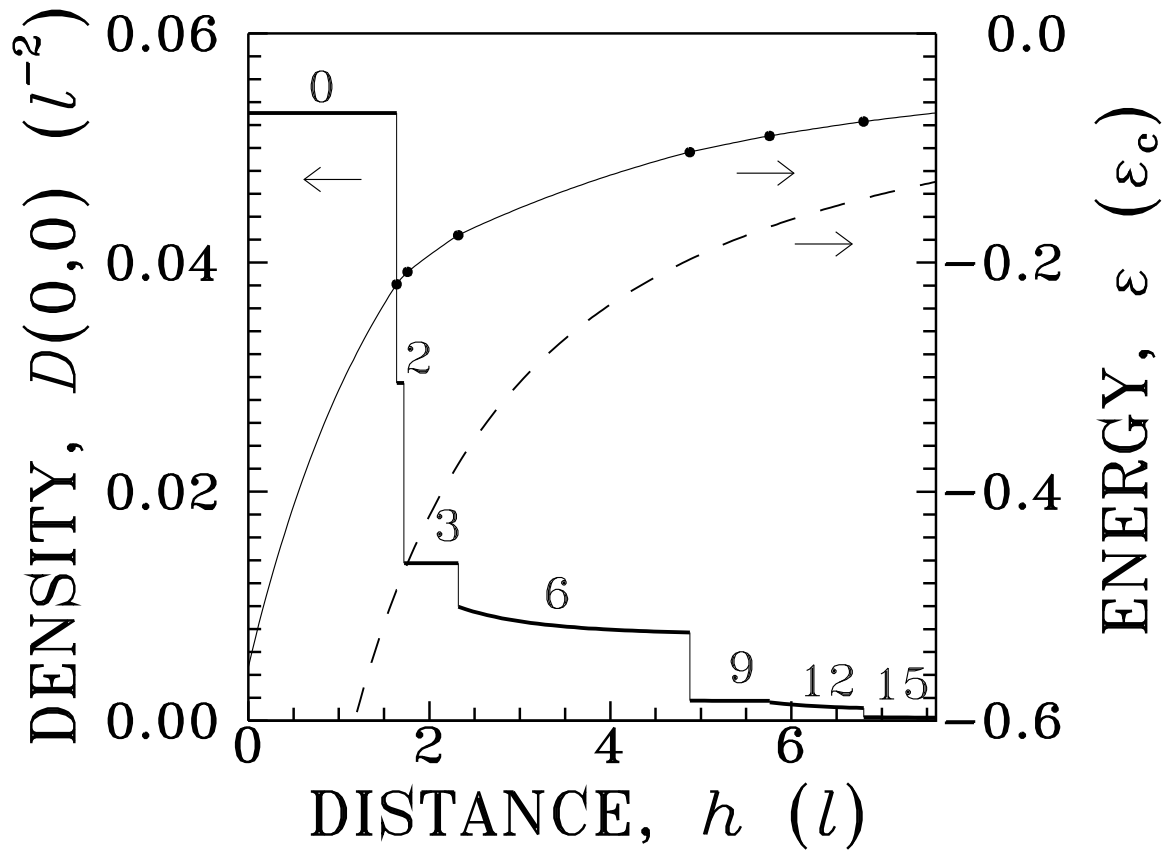}
\end{center}
\vskip 16.0truecm
 \caption{ The energy $\varepsilon(0)$ and the
 electron density $D(0,0)$ at the point
${\bf r}_{h}$ where the hole resides plotted {\it vs} $h$
for the ground state of an exciton with $k=0$. The ground state energy of an
anyon exciton is shown by a solid line; the dots on it show the
positions of the intersections between the energy levels with different
 $L$ values. For comparison the energy of a conventional magnetoexciton
$\varepsilon_{\rm me}(h)$ with $k=0$
is shown by a dashed line. Numbers near the $D(0,0)$ curve show
the $L$ values.  Only
the states with $L=3m$ reach the spectrum bottom (as an exclusion the
state $L=2$ appears as a bottom state in an extremely narrow region
of the $h$ values).
 }
\label{fig4}
\end{figure}

 Fig.~\ref{fig4} shows the energy, $\varepsilon(0)$, and the density 
 $D(0, 0)$ for bottom states as a function of $h$. 
 It is seen from the figure
 that  for $h\agt 2$ the density falls well below its critical
 value 0.05, which supports our above conjecture, Sec.~\ref{sec:Energy},
 that the AEM provides reliable results
  for $h\agt 2$. With increasing $h$ the
 angular momentum in the bottom state, $L$, also increases. It is a 
 striking feature of the data that only states with $L=3n$, $n\geq 2$,
 reach the spectrum bottom (we cannot make definite conclusions about
 the $L=3$ state since it does nor reach the bottom for $\alpha = -~ 1/3$
 anyons, Fig.~\ref{fig1}). The bottom states described by $L$-even and 
 $L$-odd polynomials alternate.
 We attribute the periodicity in $L$ to the superselection rule
 originating from the combination of the space and permutation symmetry.
 Indeed, we observe this periodicity in the semiclassical region
 where the  quantization rule includes angular integration
 between two exchange points separated by the angle $2\pi/3$
 rather than the usual $2\pi$ integration.\cite{Sivan}

 Polynomials $P_{L, M}$ with $L = 6M$ play a special role in the class of
 $L$-even polynomials. All of them obey the hard core constraint.
 Indeed, the polynomial
 $P_{6M, M}({\bar z}_{12}, {\bar z}_{23}, {\bar z}_{31})$ vanishes
 as ${\bar z}_{jl}^{L/3}$ each time when one of its arguments, 
 ${\bar z}_{jl}$, turns into zero.
  From the standpoint of the general theory,\cite{wu84} hard-core functions
 are the only "legitimate" wave functions of an anyon system.
 The exponent $L/3$ is the maximum order of the zero for a wave function
 with  the angular momentum $L$, and this maximum is achieved only for
 $P_{6M, M}$ polynomials. Therefore, Coulomb repulsion is strongly
 suppressed for these polynomials. Polynomials $WP_{6M, M}$ play
 a similar role in the class of $L$-odd polynomials. 
 It is a remarkable fact, that in the $h\agt 2$ region all bottom states
 are either $P_{6M, M}$ or $WP_{6M, M}$ polynomials.

\begin{figure}
\begin{center}
\includegraphics{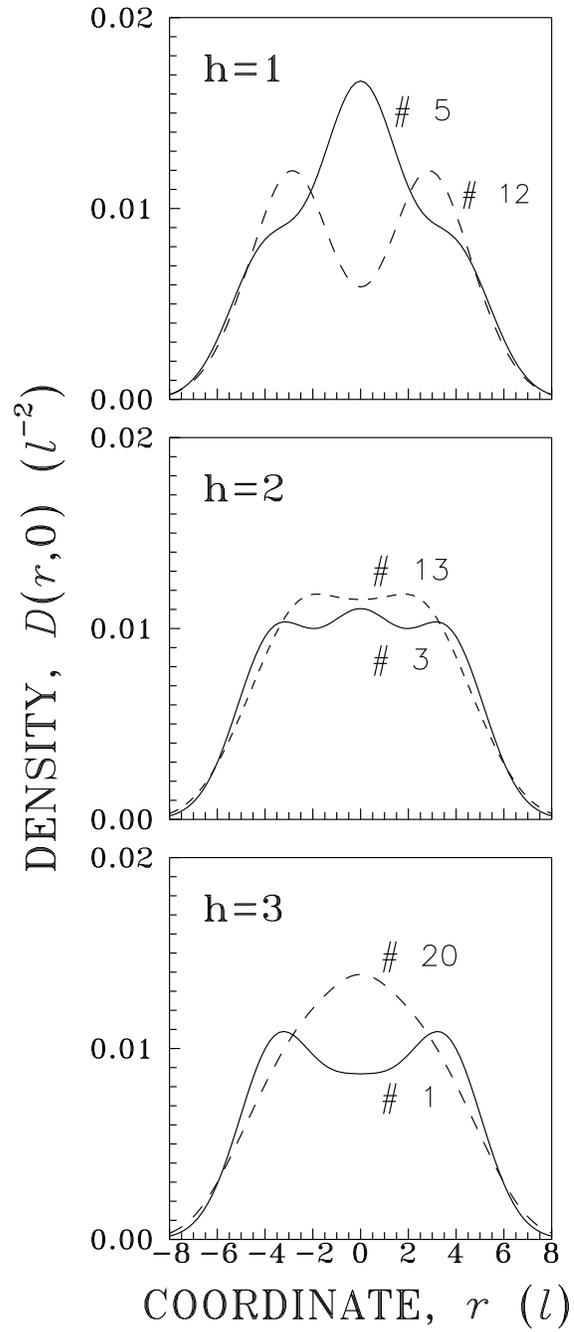}
\end{center}
\vskip 20.0truecm
 \caption{Electron density distribution $D(r,0)$ for the $L=6$ states
with $k=0$
for three values of $h$. The density in the lower energy state is
 shown by a solid line. Consecutive numbers, $\lambda$, of the energy
levels are shown near the curves.}
\label{fig5}
\end{figure}

 Using the polynomials $P_{6,M}$, $M=0$ and 1, as an example,
 we can follow the way in which hard core states move to the spectrum
 bottom when $h$ increases. For small $h$ values the low energy component
 of the $L = 6$ doublet has a pronounced maximum near $r = 0$ as it is seen
 in Figs.~\ref{fig3} and \ref{fig5}. 
 For $h<2$ the low-energy component is close to $\Psi_{6,0}$;
 for $h=0$ the overlap is 0.96. For $h \approx 2$
 wave functions of both components are 
 strongly mixed,  and they show  similar distributions of the
 density, Fig.~\ref{fig5}. For $h > 2$ the  function $\Psi_{6,1}$  wins the
 competition. For $h = 3$ it dominates in the low-energy state; 
 the overlap is 0.97. The density $D_{6M,M}(r,0)$ has a single maximum
 for each value of $M$. One can obtain a simple analytic expression
 for the position of the maximum by averaging  the density over  
 ${\bf r}_h$. This latter function, $\tilde{D}_{L,M}(r,0)$, reaches
 the maximum at $r_{L} =  \sqrt{2L}$, and the maximum of $D_{L,M}(r,0)$
 is very close to this value. More detailed information on the
 nature of the ${\bf k} = 0$ bottom states comes from the correlation functions
 which are discussed in Sec.~\ref{sec:correlation} below.
 
 For comparison, in Fig.~\ref{fig4} is also shown the $h$ dependence of the energy
 of a conventional magnetoexciton, $\varepsilon_{\rm me}(h)$,
 with the momentum ${\bf k} = 0$. In the limit $h\rightarrow 0$ this
 energy exactly coincides with the energy of an exciton in the many-electron
 system because of the hidden symmetry inherent in the problem;
 see Refs.~\onlinecite{RA} and \onlinecite{Rez93} and references therein.
 In the region of $h\ll 1$ the accuracy of the AEM is low.
 However, it increases for $h\agt 2$ when the charge fractionalization
 becomes important. In this region $\varepsilon_{\rm me}(h)$ follows
 the usual Coulomb law, $\varepsilon_{\rm me}(h) \approx - 1/h$, whereas
 for many-electron systems the dependence of the exciton binding energy
 on $h$ also is close to a Coulomb law, but the numerator is
 considerably less than 1 since the electron
 density distribution has a width about $h$. The magnetoexciton
 and AEM approaches are exact in the opposite limits. The
 results should be matched in the intermediate region at $h\approx 2$.

\begin{figure}
\begin{center}
\includegraphics{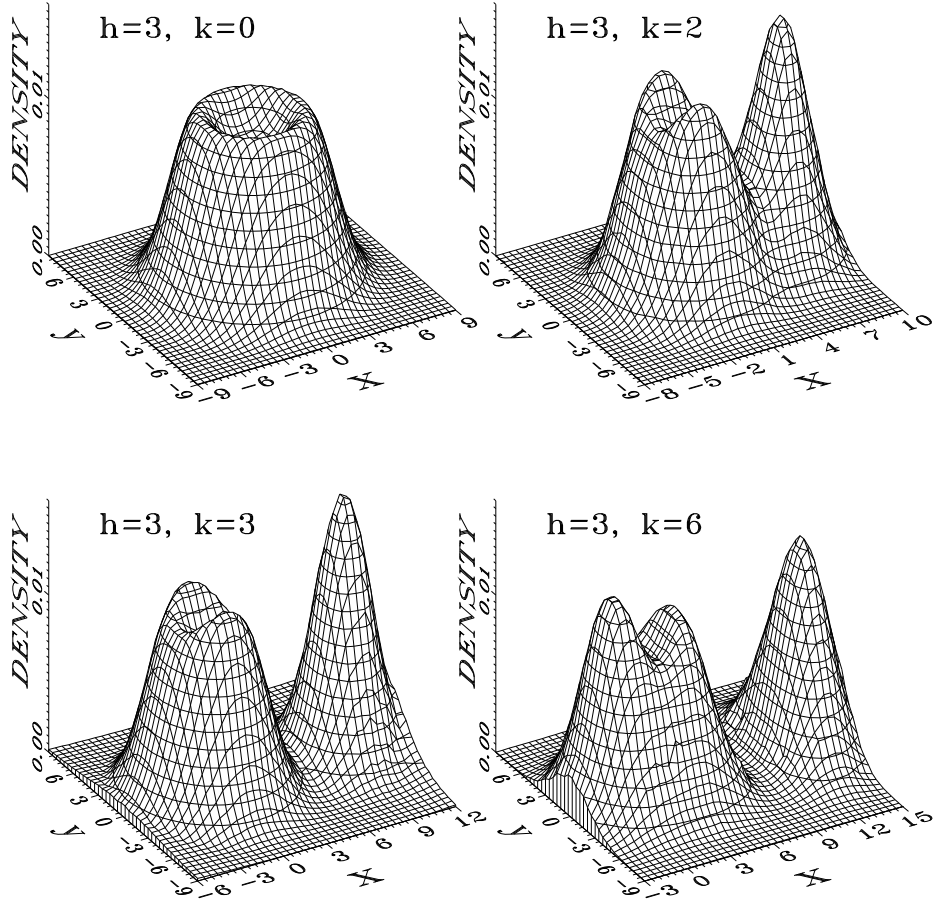}
\end{center}
\vskip 16.0truecm
 \caption{Electron density distribution in an anyon exciton for
different values of $k$. A hole is at the origin, the $x$ axis is chosen
 in the ${\bf d}$ direction. The center of the electron density distribution
is at $x=k$, y=0. The data were obtained with the 44 polynomial basis,
 $L\leq 21$; data for the 70 polynomial basis, $L\leq 27$, also show
 the change in the shape of the exciton core for large $k$ values.
 $x$,  $y$ and $h$ in units of $l$;~ $k$ in units of $l^{-1}$; density
 in units of $l^{-2}$.}
\label{fig6}
\end{figure}

 Above in this section we discussed  the electron density distribution only 
 for ${\bf k} = 0$ anyon excitons.
 The charge fractionalization manifests itself for these
 excitons in the crater shape of $D_{\lambda}(r, 0)$.
 However, the most spectacular manifestation of the charge fractionalization
 can be expected in the large $k$ region, $k\agt 1$. Indeed, the exciton
 dipole moment ${\bf d}$ differs from ${\bf k}$ only by 
 the rotation by $\pi/2$, Sec.~\ref{sec:functions}. Therefore,
 one can expect that with increasing $k$ the electron density splits into
 bundles, their charges being multiples of $1/3$.
 The splitting of the electron shell into two well separated quasiparticles
has been observed previously for a two-semion exciton.\cite{RP93} For a
three-anyon exciton the pattern are much more impressive.
 For the bottom state,  they are shown
in Fig.~\ref{fig6} for $h=3$ when the criterion of the large electron-hole 
separation is fulfilled. The distribution which is cylindrically symmetric
for $k=0$ transforms with increasing $k$ into a distribution with a
single split-off anyon ($k=2$ and 3).
 Two anyons constituting the
exciton core show a slight but distinct splitting in a perpendicular
direction. This core can be considered as an anyon ion.
 The core changes its shape with $k$ but remains stable
 in a wide range of $k$. Finally, for rather large $k$ values, it  splits in the
 ${\bf d}$ direction as it is seen in the last figure, $k=6$. 
 The asymmetric density distribution for ${\bf k}\neq 0$
 arises completely due to the admixture of 
 $L$-odd polynomials to the $L=6$ state.

 The well outlined profiles of the electron density seen in Fig.~\ref{fig6}
  may be  smeared  by the oscillatory screening
inherent in IQL's.\cite{screening} Nevertheless, the basic pattern of the
charge separation in an exciton should strongly influence the $k$
dependence of the magnetoroton-assisted recombination processes since
charge density excitations are left in a crystal afterwards.

\section{Pair correlation functions}
\label{sec:correlation}

 One can see in Fig.~\ref{fig3} that $L = 3$ state is the first state 
  which shows a crater-like shape of the density $D({\bf r}, 0)$.
 This shape indicates the existence
 of the anyon substructure of an exciton as it was
 argued in Sec.~\ref{sec:density}. It is typical of
 all bottom states with $L\geq 6$. In this section we compare 
 properties of $\Psi_{3,0}$ and  $\Psi_{6,1}$ states and show that
 despite the similarity in the shape of the density,
 they differ critically in the shape of the radial pair correlation
 function $w(r)$.

 It is convenient to use the square of the wave function $\Psi$ of 
 Eq.~(\ref{eq1.3}) averaged over the hole coordinate. Using 
 Eqs.~(\ref{eq1.3}) and (\ref{eq1.9}) and performing Gaussian integration
 over ${\bf r}_h$, one can write the following equation for the
 averaged ${\Psi}_{6M,M} $ function:
\begin{equation}
 |{\tilde \Psi}_{6M,M}({\bar z}_{12}, {\bar z}_{23}, {\bar z}_{31})|^{2}
 \propto
 (r_{12}r_{23}r_{31})^{4M}
 \exp\{- {1\over 18} (r_{12}^{2}+r_{23}^{2}+r_{31}^{2})\}~.
\label{eq7.1}
\end{equation}
 Choosing $r_{12}$, $r_{13}$, and $\psi$, where $\psi$ is the angle between
 the vectors ${\bf r}_{12}$ and ${\bf r}_{13}$, as independent variables,
 rewriting $|{\tilde \Psi}_{6M,M}|^{2}$ in terms of these variables,
 and looking for
 the maximum first over $\cos \psi$ and then over $r_{12}$ and $r_{13}$,
 one finds that the maximum of  $|{\tilde \Psi}_{6M,M}|^{2}$ is reached
 for a configuration of an equilateral triangle with
\begin{equation}
 r_{12}^{2} = r_{23}^{2} = r_{31}^{2} = 6L~.
\label{eq7.2}
\end{equation}
 In the semiclassical limit, $L\gg 1$, this result coincides with
 Eq.~(\ref{eq5.2}).

 To find the most probable configuration for the $\Psi_{3,0}$ state
 it is convenient to work in $\xi$ variables, Sec.~\ref{sec:functions},
 and perform averaging over ${\bf r}_h$.
 Simple calculation shows that
\begin{equation}
 |{\tilde \Psi}_{3,0}({\bar z}_{12}, {\bar z}_{23}, {\bar z}_{31})|^{2}
 \propto
 {\xi}^{2}_{1}{\xi}^{2}_{2}{\xi}^{2}_{3}~
 \exp\{ - {1\over 6} ({\xi}^{2}_{1} + {\xi}^{2}_{2} + {\xi}^{2}_{3})\}~.
\label{eq7.3}
\end{equation}
 The maximum of this expression under the constraint of Eq.~(\ref{eq1.7})
 can be found in the same way as for Eq.~(\ref{eq7.1}).
 Finally, one recovers an
 equilateral triangle configuration with
 $ r_{12}^{2} = r_{23}^{2} = r_{31}^{2} = 18$. This result 
 coincides with Eq.~(\ref{eq7.2}) for $L = 3$.

 Because arbitrary $L$-odd polynomials have the form $P_{L,M} = WP_{L-3,M}$,
 Eq.~(\ref{eq1.10}), and the equilateral triangle configuration is optimal
 for each of the multipliers, it is optimal also for their product, $P_{L,M}$.
 Therefore, the most probable configuration has the same shape of an
 equilateral triangle both for $L$-even and $L$-odd states.

\begin{figure}
\begin{center}
\includegraphics{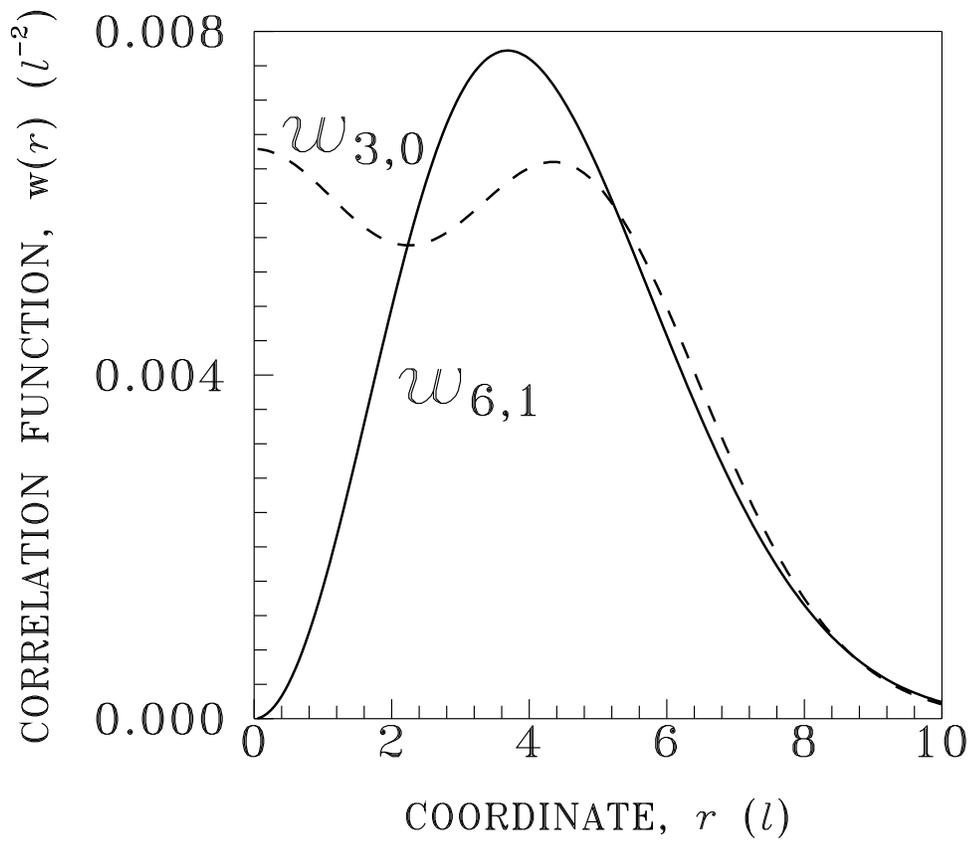}
\end{center}
\vskip 16.0truecm
 \caption{ Radial anyon pair correlation function $w(r)$ for the states
 $\Psi_{6,1}$ and $\Psi_{3,0}$; $k=0$.}
\label{fig7}
\end{figure}

 To reveal a striking difference in the properties of $\Psi_{6,1}$ and
 ${\Psi}_{3,0}$ states, one can calculate the radial pair
 correlation functions $w(r)$:
\begin{equation}
 w(r_{12}) = \int |\Psi({\bf R}, {\bbox \rho}, \{ {\bar z}_{jl}\}) |^{2}
 ~\delta({\bf r}_{12} + {\bf r}_{23} + {\bf r}_{31})
 ~d{\bf R}~ d{\bbox \rho}~ d{\bf r}_{23}~ d{\bf r}_{31} ~.
\label{eq7.4}
\end{equation}
 Substituting $|\Psi|^2$ from Eq.~(\ref{eq1.3}) results in
\begin{equation}
 w_{L,M}(r_{12}) = {1\over {(2\pi)^{2}}}
 \int d{\bf q} \int \int  d{\bf r}_{23}~ d{\bf r}_{31}
 |P_{L,M}|^{2} \exp\{ - {1\over 18}\sum_{jl} r_{jl}^{2}
 + i {\bf q} \cdot ({\bf r}_{12} + {\bf r}_{23} + {\bf r}_{31}) \}~.
\label{eq7.5}
\end{equation}
 Straightforward calculation shows that
\begin{equation}
 w_{6,1}(r)
 = {r^{2}\over 192\pi}~\{1+{1\over 2}(r/6)^{4}\}~\exp(-r^{2}/12)~,
\label{eq7.6}
\end{equation}
\begin{equation}
 w_{3,0}(r)
 = {1\over 48\pi}~\{1+{1\over 6}(r/2)^{4}\}~ \exp(-r^{2}/12)~.
\label{eq7.7}
\end{equation}
 Both functions are shown in Fig.~\ref{fig7}. The function $w_{6,1}(r)$ 
 has a hard core behavior, whereas $w_{3,0}(r)$
 does not vanish at $r=0$. On the contrary, it reaches
 its absolute maximum at this point. The
  second maximum is by the factor 0.97 lower than
the main one. Therefore, function $\Psi_{3,0}$ violates the hard core 
 constraint. Of course, all bottom-state $L$-odd polynomials with
 $L\geq 9$ show a hard core behavior.

 The above conclusion on the triangular shape of the most probable
 configuration of the  $\Psi_{3,0}$ exciton implies  existence
 of the 120$^{\circ}$ correlations in the density correlation function.

\begin{figure}
\begin{center}
\includegraphics{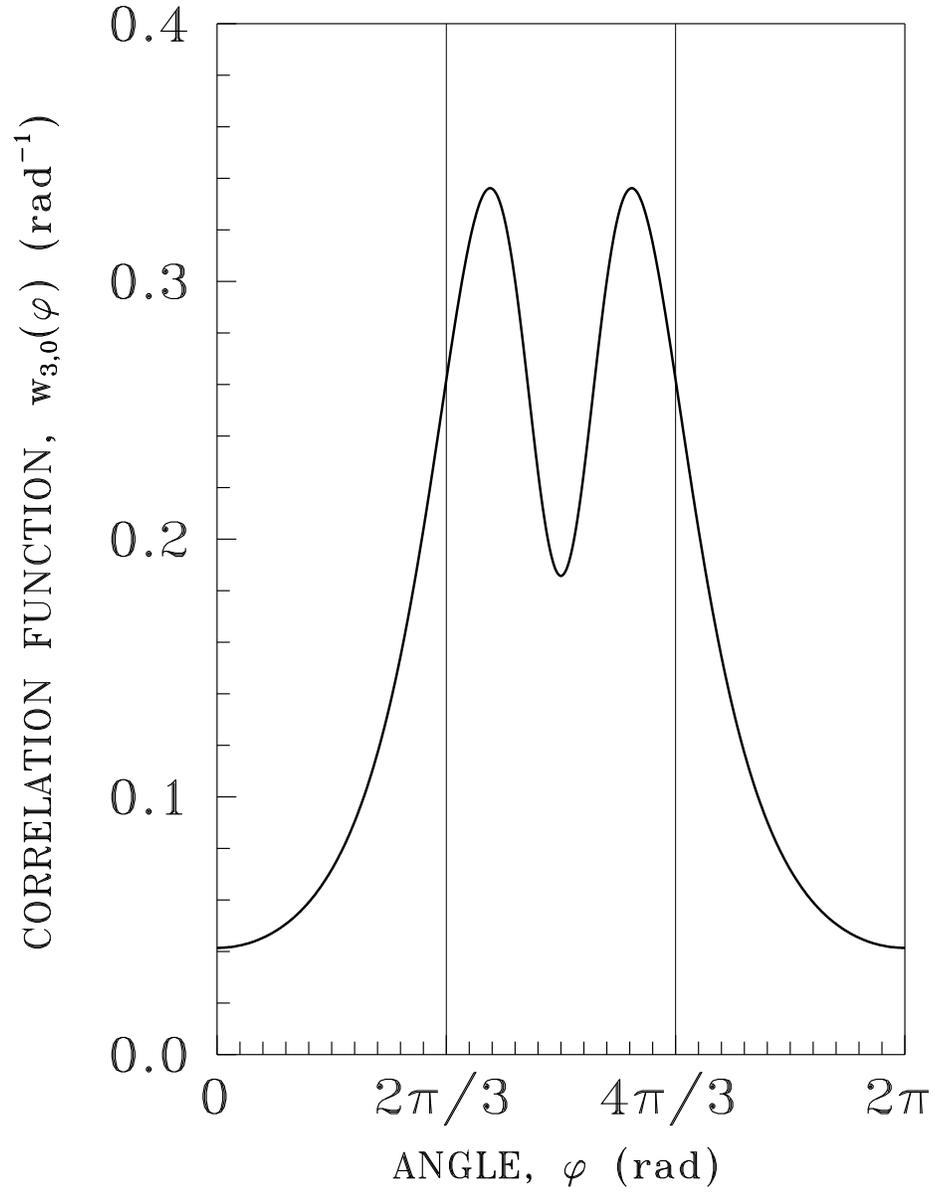}
\end{center}
\vskip 18.0truecm
 \caption{Normalized anyon pair correlation function $w_{3,0}(\varphi)$
 found in the anyon exciton model; $k=0$.}
\label{fig8}
\end{figure}

 To investigate these correlations it is convenient to work in the
 ${\bbox \xi}_{j}$ variables. 
 The two-particle correlation function $w_{3,0}({\varphi})$ depending
 on the angle $\varphi$ between vectors  ${\bbox \xi}_{1}$ and
 ${\bbox \xi}_{2}$ can be written as:
\begin{eqnarray}
 w_{3,0}(\varphi)&\propto&\int
{d{\mbox{\boldmath$\xi$}}_{1}}
{d{\mbox{\boldmath$\xi$}}_{2}}
{d{\mbox{\boldmath$\xi$}}_{3}}
|{\tilde \Psi}_{3,0}({\bbox \xi}_{1}, {\bbox \xi}_{2}, {\bbox \xi}_{3})|^2
\delta({\mbox{\boldmath$\xi$}}_{1}+
{\mbox{\boldmath$\xi$}}_{2}+{\mbox{\boldmath$\xi$}}_{3})
\delta\left(\widehat{{\mbox{\boldmath$\xi$}}_{1}
{\mbox{\boldmath$\xi$}}_{2}}-\varphi \right )
\nonumber\\
&\propto&
\int_0^{2\pi}{d\xi_1}{d\xi_2}(\xi_1 \xi_2)^3
(\xi_1^2+\xi_2^2+2\xi_1 \xi_2 \cos{\varphi})^2
\nonumber\\
&\times&
\exp\{-(\xi_1^2+\xi_2^2+\xi_1 \xi_2 \cos{\varphi})/3\}~.
\label{eq7.8}
\end{eqnarray}
 The delta function takes into account the constraint of Eq.~(\ref{eq1.7}), and
 $~\widehat{{\mbox{\boldmath$\xi$}}_{1}{\mbox{\boldmath$\xi$}}_{2}}~$
 is standing for  the angle between ${\bbox \xi}_{1}$ and ${\bbox \xi}_{2}$
 expressed in terms of the coordinates of these vectors. Integration
 over the variables  ${\bbox \xi}_{j}$ in Eq.~(\ref{eq7.8}) results
 in the averaged correlation function, the main contribution comes
 from the area
 ${\xi}_{1}^{2}\approx {\xi}_{2}^{2}\approx {\xi}_{3}^{2} \approx 6$
 where $|{\tilde \Psi}_{3,0}|^2$ reaches the maximum. The last integral can be 
 performed in polar coordinates, 
$~\xi_1=\xi \cos(\theta/2)~$, $~\xi_2=\xi \sin(\theta/2)~$,
$~0\leq\theta\leq\pi~$, and the normalized function $w_{3,0}(\varphi)$
 takes the form:
\begin{equation}
 w_{3,0}(\varphi) = {81 \over 8\pi} \int_0^{\pi}{d\theta}~
\frac{\sin^{3}{\theta}(1+\sin{\theta}\cos{\varphi})}
{(2+\sin{\theta}\cos{\varphi})^5}~.
\label{eq7.9}
\end{equation}
 The integration
 over $\theta$ can be performed analytically, however, the final
 expression is rather cumbersome. The result is plotted in Fig.~\ref{fig8}.
 Two distinct maxima in the vicinity of $2\pi/3$ and $4\pi/3$
 reveal 120$^{\circ}$ correlations in the $\Psi_{3,0}$ state.
 The maxima are shifted from the angle $2\pi /3$ since the function
 $\Psi_{3,0}$ vanishes for the collinear configuration of anyons,
 ${\bbox \xi}_{1} = - {\bbox \xi}_{2}, ~ {\bbox \xi}_{3} = 0$.

\begin{figure}
\begin{center}
\includegraphics{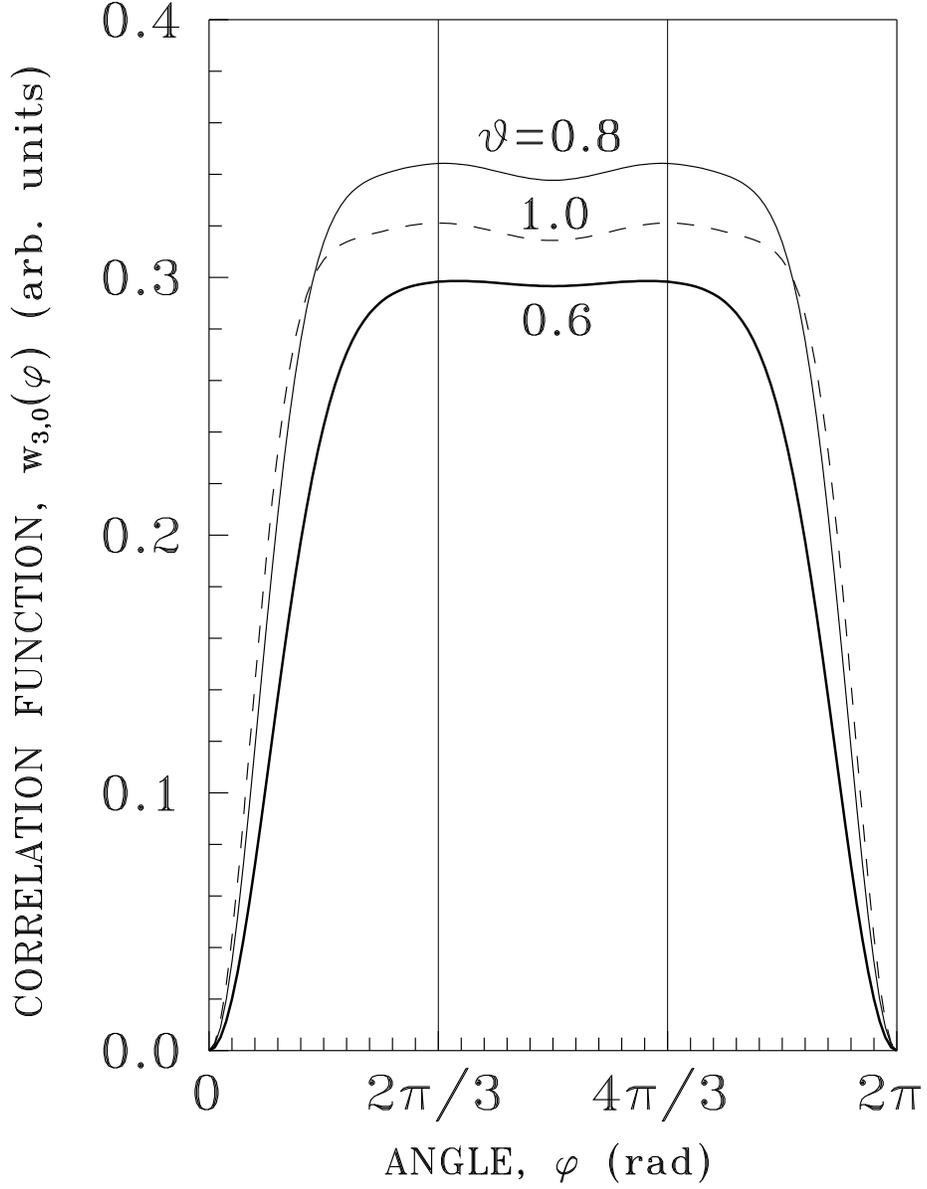}
\end{center}
\vskip 18.0truecm
 \caption{ Electron density correlation function
 $w_{3,0}^{dd}(\vartheta, \varphi)$ for the $\nu = 2/3$
 IQL with a single extra electron ($N=15$, the flux $2S=21$). 
 The correlation function
 is plotted as a function of azimuth $\varphi$ for three values of
 the polar angle $\vartheta$ = 0.6,~0.8, and 1.0.
 $\vartheta = 0.8$ is close to the
 maximum of the electron density. Correlations between quasielectrons
 manifest themselves in the maxima near $\varphi = 2\pi /3$ and $4\pi /3$.
 (Apalkov and Rashba\protect\cite{ARunpubl})}
\label{fig9}
\end{figure}

 Apalkov and Rashba\cite{ARunpubl} have found the density-density
 correlation function, $w_{3,0}^{dd}(\varphi )$,
 for the $\nu = 2/3$ IQL with a single extra electron.
 Calculations were performed in the spherical geometry for the
 $ (L_{QP})_{\rm max} - L_{QP} = 3$, $(L_{QP})_{z} = L_{QP}$
 quantum state. 
 The quantity $ (L_{QP})_{\rm max} -  L_{QP}$,
 which is the difference between the three-quasiparticle angular
 momentum in the spherical geometry, $L_{QP}$,  and the maximum value of
 this momentum, should be compared to
 the exciton angular momentum $L$.\cite{APR,AR2/3}
 It is convenient to introduce mean values
\begin{equation}
 \rho_{1}(\vartheta) = \int |\Psi({\bbox \omega}, {\bbox \omega}_{2},
 \ldots {\bbox \omega}_{N})|^{2} ~d{\bbox \omega}_{2}
 \ldots d{\bbox \omega}_{N}~
 \label{eq7.10}
\end{equation}
and
\begin{equation}
 \rho_{2}({\bbox \omega}, {\bbox \omega}^{'}) =
 \int |\Psi({\bbox \omega}, {\bbox \omega}^{'}, {\bbox \omega}_{3}
  \ldots {\bbox \omega}_{N})|^{2} 
 ~d{\bbox \omega}_{3} \ldots ~d{\bbox \omega}_{N} ~.
\label{eq7.11}
\end{equation}
 Here ${\bbox \omega}_{j}(\vartheta, \varphi)$ are unit vectors designating the
 positions of the electrons on the sphere. Electron density depends only
 on the polar angle $\vartheta$ and equals 
 $n_{1}(\vartheta) = N\rho_{1}(\vartheta)$, where
 $N$ is the number of electrons.
 If one introduces the deviation,
 $\Delta n({\bbox \omega}) = n({\bbox \omega}) - n_{1}(\vartheta)$,
 of the density from its mean value,
 the density-density correlation function can be written as
\begin{equation}
 w_{3,0}^{dd}(\vartheta, \varphi - \varphi^{'}) =~
<\Delta n(\vartheta, \varphi) ~\Delta n(\vartheta, \varphi^{'})> ~=
 {1\over 2}N(N-1)~ \rho_{2}(\vartheta, \varphi - \varphi^{'}) -
 n_{1}^{2}(\vartheta)~.
\label{eq7.12}
\end{equation}
 In Eq.~(\ref{eq7.12}) the polar angles of the vectors ${\bbox \omega}$ and 
 ${\bbox \omega}^{'}$  are chosen equal, $\vartheta = \vartheta^{'}$.
  Therefore, the correlation function, $w_{3,0}^{dd}(\vartheta, \varphi)$, 
  is the function of the azimuth $\varphi$ and
 depends on $\vartheta$ as a parameter.
 In Eq.~(\ref{eq7.12}) the singular term 
 proportional to $\delta({\bbox \omega} - {\bbox \omega}^{'})$ 
 is omitted, as usual, since
 it makes no contribution to $w_{3,0}^{dd}$ for
 ${\bbox \omega}\neq {\bbox \omega}^{'}$.
 The results of calculations are
 presented in Fig.~\ref{fig9} for three values of $\vartheta$.
 The value of $\vartheta = 0.8$
 is close to the maximum of $n_{1}(\vartheta)$.

 Since the anyon exciton wave function $\Psi_{3,0}$ does not depend on $h$,
 the data of Figs.~\ref{fig8} and \ref{fig9} can be compared.
 There is a striking similarity between them. All three
curves in Fig.~\ref{fig9} show flat but distinct maxima near the same values
 of the argument,
 $\varphi = 2\pi/3$ and $4\pi/3$.
 These maxima are much flatter than the maxima in Fig.~\ref{fig9}. 
 This difference can be attributed to the smearing originating from 
 the quasielectron form-factors. Nevertheless, 
  there is no doubt that it is the three quasielectron substructure
  of the many electron state $L = 3$ which manifests itself in
  120$^{\circ}$ correlations. 

 Therefore, the state $\Psi_{3,0}$ shows properties intermediate between an
 anyon exciton and tight exciton. Unfortunately, reliable calculation of
 $w^{dd}(\vartheta, \varphi - \varphi^{'})$ can not be performed for the
 $(L_{QE})_{\rm max}-L_{QE} = 6$ states since their size exceeds the
 accessible sphere size.

 \section{Comparison with finite size calculations}
\label{sec:comparison}

 In this section we make a comparison of the AEM
 with  computational results for excitons
 in the spherical geometry. Three-anyon
 excitons are expected to appear in two charge conjugate IQL's,
 $\nu = 1/3$ and $\nu = 2/3$. Their properties in the semiclassical
 region, $h\gg 1$ and $L\gg 1$, should be identical. 
 For intermediate $h$ and $L$ values, excitons of these two liquids are expected
  to show rather different properties. The data obtained
 by finite-size computations\cite{APR,AR2/3} substantiate these expectations. 

 There are different reasons why $\nu = 1/3$ and $\nu = 2/3$ excitons
 are expected to have different properties in the intermediate region.
 A simple electrostatic reason is a different
 nonlinear screening of the unit positive charge by these two IQL's. 
 The effect of the screening on the exciton energy spectrum was discussed
 in Ref.~\onlinecite{TRE}. However, there are also different mechanisms
 resulting in the difference in the properties of these excitons. They are
 related to rather different form-factors and different statistical
 properties of the quasielectrons of the $\nu = 1/3$
 and $\nu = 2/3$ IQL's. Neglecting spin effects, we can consider electrons
 as spin-less fermions. Then, because of the charge symmetry, the properties
 of the quasielectrons of the $\nu = 2/3$ IQL are identical to the properties
 of the quasiholes of the $\nu = 1/3$ IQL. Therefore, we will start with a
 comparative study of quasielectrons and quasiholes of the $\nu = 1/3$ IQL.

  The form-factors of charged particles of the $\nu = 1/3$ IQL
 were investigated in a number of papers. The data are summarized
 in Refs.~\onlinecite{CP} - \onlinecite{GH94}.
 It is known that a quasihole has a narrow
 profile with a radius up to two magnetic lengths. The density 
 decreases away from the center of a quasihole nearly monotonically.
 On the contrary, a quasielectron has a pronounced density dip at the center, 
 the density maximum at about two  magnetic lengths,
 and the radius of about four
 magnetic lengths.
 Therefore, the model of point anyons developed above matches much better
 the excitons of the $\nu = 2/3$ IQL than the excitons of the $\nu = 1/3$
 IQL. 

 Another aspect of the problem is related to statistical properties of
 anyons. In the AEM the effect of the fractional
 permutational
 statistics of anyons was taken into account by including the factor
 $\prod_{jl} ({\bar z}_{jl})^{\alpha}$ into the Halperin
 pseudo-wave-function
  $\Psi_{L,{\bf k}}$~, Eq.~(\ref{eq1.3}). It was shown
 in Sec.~\ref{sec:Energy} that the
 effect of this factor can be neglected for $h\agt 2$,
 i.e., in the region where
 the AEM is expected to be valid.
 However, one should also take into account the nontrivial populational
 statistics of anyons. This can be done using the theory
 of composite fermions\cite{Jain} and the approach to the
 dimensionality of the quasiparticle space
 based on exclusion statistics.\cite{Hal91,WuExcl} Bosonic
 Haldane dimension, $d^B_{QP}$, is an effective number of the
 single-quasiparticle states defined in such a way that the usual
 Bose distribution
\[
 W_{B}(d^{B}_{QP}, N_{QP}) =
 (d^{B}_{QP} + N_{QP} - 1)!/(d^{B}_{QP} - 1)! N_{QP}!
\]
 results in the correct number of states in the Hilbert space
 of $N_{QP}$ quasiparticles,
 $W(N_{QP}) = W_{B}(d^{B}_{QP}, N_{QP})$.
 The number of quasiparticle states, $W(N_{QP})$,
  can be found by counting the number of states
 in the Hilbert space of composite fermions.\cite{DJ92,Jain}
 If quasiparticles obey Bose statistics, $d^{B}_{QP}$ does not depend
 on $N_{QP}$. By counting the composite fermion number-of-states
 in the spherical geometry, it was shown in Ref.~\onlinecite{AR2/3}  
 that for $\nu = 1/m$, $m$ is an integer, the bosonic dimension of
quasiholes equals
\begin{equation}
 d^{B}_{QH} = N + 1~,
\label{eq8.1}
\end{equation}
and the bosonic dimension of quasielectrons equals
\begin{equation}
 d^{B}_{QE} = (N + 1) - 2(N_{QE} -1)~.
\label{eq8.2}
\end{equation}
 Here $N$ is the number of electrons. These equations are consistent
 with the diagonal coefficients of the exclusion statistics of quasiholes
 and quasielectrons ($1/m$ and $2 - 1/m$, respectively), found by different
 authors.\cite{JC94,Yang,SWY96}

 Equations (\ref{eq8.1}) and (\ref{eq8.2}) indicate that 
 quasiholes show a bosonic behavior, while quasielectrons are subject
 of the constraint which is even more restrictive than the constraint
 imposed by Fermi statistics. These conclusions were supported\cite{AR2/3}
 by counting the maximum value of the angular momentum,
 $(L_{QP})_{\rm max}$~, for the system 
 of $N_{QP}$  quasiparticles. It is equal to
\begin{equation}
 (L_{QH})_{\rm max} = N_{QH}N/2
\label{eq8.3}
\end{equation}
 for $N_{QH}$ quasiholes, and to
\begin{equation}
 (L_{QE})_{\rm max} = N_{QE}N/2 - N_{QE}(N_{QE} - 1)
\label{eq8.4}
\end{equation}
for $N_{QE}$ quasielectrons. Since $N/2$ is the angular momentum of a single
quasiparticle,\cite{Hal83} Eq.~(\ref{eq8.3}) confirms the bosonic
 behavior of quasiholes, whereas Eq.~(\ref{eq8.4}) confirms the existence
 of the restriction on the population of single-quasiparticle states by
 quasielectrons. He {\it et al.}\cite{HXZ} were  the first to 
 discover this restriction by means of numerical calculations
 and to propose Eqs.~(\ref{eq8.1})
 and (\ref{eq8.2}). They attributed the restriction to the hard core
 constraint for quasielectrons having a dynamical (short-range repulsion)
 rather than statistical origin. 

 For a macroscopic system, $N\gg 1$, equations (\ref{eq8.1}) and (\ref{eq8.2})
 give coinciding results, $d^{B}_{QE}\approx d^{B}_{QH} \approx N$,
 in the dilute limit,
 $N_{QP}\ll N$. However, in an exciton the quasiparticles are confined
 inside the volume about $\pi r^2$, where $r$ is the exciton radius. Therefore,
  the second term of  Eq.~(\ref{eq8.2}) which
 signifies the deviation from the bosonic behavior of quasielectrons,
 can be of importance.

 Quasielectrons of the $\nu = 2/3$ IQL 
 are described by Eqs.~(\ref{eq8.1})
 and (\ref{eq8.3}). Therefore, the population of single-quasiparticle
 states obeys the Bose statistics, and Eq.~(\ref{eq1.3}) for wave functions
 is absolutely adequate since it includes polynomials $P_{L,M}$
 symmetrical in variables ${\bar z}_{j}$. For this reason, and
 also taking into account narrow form-factors of quasielectrons of the
 $\nu = 2/3$ IQL, one can conclude that the $\nu = 2/3$ IQL is the best
 candidate for comparison with the AEM. We
 believe, that the criterion $h\agt 2$, established in Sec.~\ref{sec:density}
 by evaluating the density $D_{\lambda}(r, 0)$, is applicable to the
 $\nu = 2/3$ IQL.
 
 The situation is more involved for the $\nu = 1/3$ IQL. Quasielectrons
 in this case are described by Eqs.~(\ref{eq8.2}) and (\ref{eq8.4}).
 One can apply Eq.~(\ref{eq8.2}) to the area inside an exciton and
 order that $d^{B}_{QE}\geq N_{QE} = 3$, which results in $N\geq 6$.
 The number of electrons inside an exciton can be evaluated as
 $N\approx \pi r^{2}/2\pi$. Evaluating the exciton radius $r$ as 
 $r\approx r_{12}/\sqrt{3}$ and
 using the classical equation (\ref{eq5.1}),
 one comes to the criterion $h\agt 5$. 
 A wide quasielectron form-factor imposes a similar restriction on $h$. Since
 the reliable finite-size computations can be performed
  only for $h\alt 2$, the
 prospects for a quantitative comparison of the results obtained by
 both approaches seem  less favorable for the $\nu = 1/3$ IQL 
 than for the $\nu = 2/3$ IQL. However, we feel that the above rigid
 criterion relaxes considerably when a qualitative description
 of the spectrum-bottom states is concerned. 

 We are now in position to compare the basic  results of the AEM
 with the computational results for finite-size systems. The basic
 statement of the theory of anyon excitons, that the charge fractionalization
 results in a multiple-branch energy spectra of excitons,\cite{RP93,MPL}
  was confirmed by finite-size
 computations performed for both 
 $\nu = 1/3$ and $\nu = 2/3$ IQL's. Actually, it
 provides the basic idea for representing the energy spectra
 obtained in the spherical geometry for a discrete set of 
 $L$ values in the form of exciton branches
 $\varepsilon_{\lambda}({\bf k})$.
 
 We begin with summarizing  some results obtained in Ref.~\onlinecite{AR2/3}
 by finite-size calculations
 for excitons of the $\nu = 2/3$ IQL  and compare
 them with the results obtained in this paper in the framework
 of the AEM. 
 Classification of excitons in terms of tight and anyon excitons is used,
 cf. Sec.~\ref{sec:introduction}.

 1. {\it Number of exciton species}. Anyon exciton is a bound state
 of a three-quasielectron complex from the low-energy (anyon)
  sector\cite{HXZ,DJ92} and a hole. The angular momentum of the exciton,
 $L$, is equal to $L^{*} = (L_{QE})_{\rm max}-L_{QE}$, where $L_{QE}$
 is the angular momentum of the complex, and $(L_{QE})_{\rm max}$ is
 the maximum value of this momentum which can be found from Eq.~(\ref{eq8.3}).
 Therefore,
 the number of exciton species is equal to the number of the
 three-quasielectron complexes in the low-energy sector, and $L^{*}$
 should be compared to the exciton angular momentum $L$ of the AEM.
 Only a single three-quasielectron complex exists
 if the angular momentum $L^{*}$ equals $L^{*}$ =  0, ~2, ~3, ~4, ~5, or 7.
 The complex $L^{*} = 1$ is absent because
 composite fermions obey Fermi statistics. For $L^{*}\geq 6$ the number of
 $L$-even states increases by one (see Fig.~1 in Ref.~\onlinecite{AR2/3}).

 These properties are in a complete agreement with the
 classification of the polynomials $P_{L,M}$ of Sec.~\ref{sec:functions},
 Eqs.~(\ref{eq1.9}) and (\ref{eq1.10}). The $L = 1$ exciton is absent
 because of the constraint of Eq.~(\ref{eq1.2}).

 2. {\it L} = 0 {\it branch}. Despite the fact that the $L=0$ exciton
 appears, according to its quantum number, as the first state in the
 series of anyon excitons, it possesses rather special properties. 
 This exciton originates from the $L^{*}=0$
 three-quasielectron complex which is
 quite dense, and the energy of this complex is high. It is
 nearly the same as the energy of some states from the next sector. These data
 imply that the $L=0$ anyon exciton actually merges with the $L=0$
 tight exciton. Because of these arguments, the $L = 0$ exciton was assigned in
 Ref.~\onlinecite{AR2/3} as a tight exciton rather than an anyon one.
 This assignment is supported by an independent argument. $L=0$
 excitons of the $\nu=2/3$   and $\nu=1/3$ IQL's show nearly identical
 properties, whereas the anyon exciton assignment of the latter entity
 is excluded by the symmetry arguments based on the composite
 fermion theory.\cite{APR}

 The shape of the $L = 0$ exciton density distribution of Fig.~\ref{fig3} is in
 agreement with this assignment. The magnitude of $D_{0}(0, 0)$ exceeds
 the maximum density compatible with the AEM; see the discussion at the
 beginning of Sec.~\ref{sec:density}.

 3.  {\it Electron density}. For a system with a single extra electron
 against a background of the $\nu = 2/3$ IQL,
 the electron density has a pronounced maximum at $r = 0$ if 
 $L^{*} = 0, ~2,$ or 4. The $L^{*} = 3$
 state is the first state with a crater-like
 electron density distribution having a density dip at $r = 0$
 (see Fig.~2 in Ref.~\onlinecite{AR2/3}). The crater shape of the density
 indicates the charge fractionalization.

 These data are in a qualitative  agreement with the electron density
 distributions of Fig.~\ref{fig3} for excitons 
 having angular momenta $L$ = 0,~ 2, ~3, and 4.
 The general shape of the curves is the same, but there are
  differences in the magnitudes of the densities at $r = 0$.
 Since wave functions of
 these excitons do not depend on $h$, electron densities 
 for the exciton and three-quasielectron states with 
 the angular momenta $L=L^{*}$, respectively, can be  compared.

 4. {\it Bottom states}. $L=0$ exciton remains the bottom state in the
 whole area $h\alt 2$. $L = 2$, $L = 3$, and $L =6$ excitons move down to the
 spectrum bottom, but they start competing with the $L = 0$ exciton only for
 $h\approx 2.5$ when the accuracy of finite-size calculations becomes
 ambiguous. $L = 4$ exciton is also seen in the low-energy part of the
 spectrum but never reaches the spectrum bottom.\cite{AR2/3,ARunpubl}

 These data are in agreement with Fig.~\ref{fig4} where the sequence of the 
 first bottom states includes $L=0,~2,~,3$ and 6.

 5. {\it Charge fractionalization: density correlation function}.
 It was shown in Sec.~\ref{sec:correlation} that 
 $w_{3,0}({\varphi})$ 
  reveals  120$^{\circ}$ correlations
 both for the many-electron and AEM wave functions. These
 correlations signal the charge fractionalization.

 6. {\it Intrinsic angular momenta of anyons}. There is a convincing
 one-to-one correspondence between the excitons of  many electron
 systems having angular momenta $L\leq 6$ in the spherical geometry 
 and the excitons
 of the AEM with the same values of the $z$ projections  of the
 angular momentum. Therefore, our data provide no indication of the
 existence of the intrinsic angular momenta 
 of anyons.\cite{LeeWen}
 On the contrary, our data are 
 in agreement with the recent conjecture on the absence of anyon spins
 in the plane limit.\cite{Einarsson}

 The AEM predicts identical exciton spectra for the
 $\nu = 2/3$ and $\nu = 1/3$ IQL's. However, finite-size
 calculations\cite{APR,AR2/3} result in a rather different symmetry
 of the low-energy exciton states for $h\alt 2$. The
 exact classification of the exciton states based on the composite
 fermion theory shows that only $L=3$ and $L\geq 5$ anyon excitons
 can exist in the $\nu =1/3$ IQL.\cite{APR} This conclusion is supported
 by numerical data. Therefore, a challenging question arises: Why
  $L=2$ and $L=4$ excitons of the AEM are missing from the many-electron
 spectra of the $\nu =1/3$ IQL? We argue, that these excitons are
 excluded because of their small bosonic dimension
 which cannot accommodate three quasielectrons.

 In what follows we compare the results of the
 finite-size computations of Ref.~\onlinecite{APR} for the $\nu = 1/3$ IQL
 with the AEM data.
  
 1. {\it Excluded states}. Finite-size calculations
 and composite fermion theory  show that anyon
 excitons of the $\nu = 1/3$ IQL can only posses angular momenta $L=3$
 and $L\geq 5$. All excitons with $L$ = 0,~1,~2, and 4 can
 only appear as  tight excitons. $L=0$ tight exciton forms
 the spectrum bottom for $h\alt 2$, but there is no low-energy
 $L=2$ and $L=4$ excitons.

 It seems probable that there can exist only one, $L=0$, tight exciton
 near the spectrum bottom. Therefore, it is necessary to understand why 
 $L=2$ and $L=4$ excitons of the AEM, which possess low energies, do
 not appear as anyon excitons in many-electron systems. The
 criterion $d^{B}_{QE}\geq N_{QE}$ results in the minimum
 anyon-exciton radius of $r_{\rm min}\approx 2\sqrt{3}$. 
 It is seen in Fig.~\ref{fig3}
 that the criterion $r\agt r_{\rm min}$ is violated for
 $L=2$ and $L=4$ excitons. Therefore, we
 attribute the exclusion of the $L=2$ and $L=4$ exciton states 
 from the many-electron spectrum to the reduction of the bosonic
 dimension of the quasielectron space because of the second term of
 Eq.~(\ref{eq8.2}).

 2. {\it Bottom states.} Tight $L = 0$ exciton remains the bottom state
 up to $h\approx 2$ when the $L = 3$ anyon exciton reaches the bottom.
 $L=5$ and $L=6$ excitons 
 start to compete with the $L=0$ exciton only for $h\approx 3$
 when the accuracy of finite-size calculations is low.

 The data are in agreement with Fig.~\ref{fig4}.

 3. {\it Electron density distribution for} ${\bf k} = 0$ {excitons}.
 Tight excitons have narrow electron density distributions.
 The density of the
 $L = 0$ exciton is sharply peaked at $r = 0$. $L = 3$ and $L = 5$
 anyon excitons show pronounced density dips at $r = 0$ and two
 maxima of the density. It was hypothesized\cite{APR} 
 that the $r = 0$ dip originates from the dip in the quasielectron
 form-factor, whereas the two-maxima shape indicates the existence
 of a two-anyon core and one split-off anyon. It is remarkable that
 the density distribution  in the $L = 3$ and $L = 5$ excitons only
 weakly depends on $h$.

 Weak $h$ dependence of the density distribution for $L = 3$ and $L = 5$ 
 excitons is in agreement with the existence of a single polynomial
 $P_{L,M}$ for $L\leq 6$, Sec.~\ref{sec:functions}. Splitting of the exciton
 shell of an ${\bf k} = 0$ exciton can not be described within the framework
 of the point-anyon AEM.

 4. {\it Dependence  of the density distribution on} ${\bf k}$.
 When $k$ increases, the $r = 0$ dip in the electron density of the
 $L = 3$ exciton transforms into a narrow maximum, and the density
 distribution acquires a three maxima shape. It was proposed\cite{APR}
 that in this region of $k$ values an exciton consists of a single-anyon core
 and a two-anyon shell.

 Splitting off of anyons from the core with increasing $k$ is in a
 qualitative agreement with Fig.~\ref{fig6}.

\section{Conclusion}
\label{sec:conclusion}

 The model of anyon exciton developed and solved in the paper includes
 three quasielectrons (anyons) and one hole. It is applicable to exciton
 spectra of two charge conjugate IQL's, $\nu =1/3$ and $\nu =2/3$,
 and is exact in the limit
 of a large separation between electron and hole confinement planes,
 $h\gg l$.

Anyon excitons possess multiple branch energy spectra,
 $\varepsilon_{\lambda}({\bf k})$. An exciton is described by a 2D
 momentum $\bf k$ and two internal quantum numbers $\lambda$
 which numerate exciton branches. One of these quantum numbers
 can be chosen as the exciton angular momentum, $L$, in the ${\bf k}=0$ state.
 
 A full set of basis functions was chosen with a proper account of the
 magnetic translational symmetry and permutational symmetry.
 The functions include
 two types of polynomials symmetric in anyon permutations,
 one of them, apparently, was considered
 for the first time. Analytic expressions for all matrix elements were
 derived. As a result, exact expressions for the energy spectrum of
 a four-particle system were found for ${\bf k}=0$. All exciton
 states with even angular momenta $L < 6$ and odd momenta $L < 9$
 are non-degenerate, and their wave functions are completely determined
 by symmetry requirements. This property manifests itself in finite-size data
 in a weak $h$ dependence of the electron density distribution.
 Properties of ${\bf k}\neq 0$ excitons
 were investigated by numerical solution of the Schroedinger equation.
 
 The AEM is not applicable for small $h$ values, $h\alt 2l$. However,
the analysis of the internal criteria of the AEM and comparison with the 
 finite-size data show that it gives satisfactory results for $h\approx 2l$,
 especially for excitons of the $\nu =2/3$ IQL.
 Finite-size computations which are reliable for $h\alt 2l$, and the AEM,
 whose accuracy increases with $h$, constitute two supplementary approaches  
  which match at $h\approx 2l$.
 According to the AEM, exciton states with the
 angular momenta obeying the superselection rule $L=3m$,
 where $m\geq 2$ is an integer,
 form the sequence of the bottom states for $h\agt 2l$.
 All these states are hard core states, i.e., their
 wave functions turn into zero if any two of the anyon coordinates
 coincide. The equilateral anyon configurations are the most probable ones.
 Angular momenta $L$ of the bottom states increase as $h^2$,
 and the size of these states
 as $h$. The tight $L=0$ state is the bottom state for $h\alt 2l$,
 whereas $L=3$ state, which possesses the properties of both anyon
 and tight excitons, can reach the bottom for $h\approx 2l$.

 Anyon substructure of excitons manifests itself in their different properties.
 First, the multiple branch energy spectrum originates because
 of the existence of the internal degrees of freedom. Second, 
 for ${\bf k} = 0$ excitons a
 crater-like shape of the electron density distribution 
 with a pronounced dip near the hole signals the existence
 of the charge fractionalization. Third, 120$^{\circ}$ correlations in
 the electron density were found both in the framework of the AEM and in
 finite-size data. Fourth, for ${\bf k}\neq 0$ excitons
 a spectacular splitting of the electron
 density into bunches provides a direct
 manifestation of the fractional charge substructure. 

 AEM results in identical low-energy spectra for the excitons of the
 $\nu =1/3$  and $\nu =2/3$ IQL's. This result is definitely correct
 in the $h\gg l$ limit. However, finite-size data suggest that in the
 intermediate region, $h\approx l$, the $\nu =1/3$ IQL possesses less
 anyon exciton branches than the $\nu =2/3$ IQL. Following 
 Ref.~\onlinecite{AR2/3}, we attribute this phenomenon to the 
 difference in the populational statistics of the quasielectrons of
 the charge conjugate IQL's. Anyon statistics enters into the 
 theory in two ways. First, through the fractional-power factors
 in the wave function; they ensure correct interchanging statistics.
 Second, through the populational statistics. We have shown that
 the first mechanism can be neglected for $h\agt 2l$, whereas the
 second one is of a critical importance in the intermediate region.
 It follows from the composite fermion theory that quasielectrons of the 
 $\nu =2/3$ IQL obey bosonic populational statistics. As a result,
 quantum numbers of anyon excitons found from the AEM and from the
 finite-size data exactly coincide. On the contrary, the bosonic
 dimension, $d^{B}_{QE}$, of the quasielectron space
 of the $\nu =1/3$ IQL rapidly decreases with the number
 of quasielectrons, $N_{QE}$.
 The condition $d^{B}_{QE}\geq N_{QE} $, written for the area
 about the exciton size, eliminates the exciton states
 of the AEM with the angular momenta $L=2$ and $L=4$ 
 from the many-electron spectrum 
  and brings in agreement the AEM and
  many-electron data. Therefore, for $h\approx 2l$ the AEM
supplemented with anyon statistics arguments matches the many-electron data
 for both IQL's and sheds light on the origin of the difference
 in their exciton spectra.
 The above arguments are rather general and can be applied
 to different problems of the theory of the bound states of several anyons.
 
 Comparison of the AEM and finite-size data for $\nu =2/3$ IQL provides
 no indication of the existence of the intrinsic angular momenta (spins)
 of anyons.

 In conclusion, the anyon exciton model unveils the general pattern
 of the exciton spectra of IQL's. It predicts the properties of excitons
 in the large $h$ limit, $h\gg l$, and is in agreement with finite-size
 data in the intermediate region, $h\approx 2l$. The theory suggests that
 the region $h\agt 2l$ is most favorable for investigating the anyon
 substructure of excitons in optical experiments.

\acknowledgments

 We are grateful to  A. L. Efros and Y.-S. Wu for discussions and comments,
 and to E. V. Tsiper for suggestions related to $L$-odd
 polynomials. E.I.R. acknowledges the support of the San Diego
 Supercomputer Center where finite-size
 computations were performed. We also acknowledge the support of the QUEST
 of the UCSB by subagreement KK3017.

\newpage

\newpage

\newpage

\newpage

\newpage

\newpage

\newpage

\newpage

\newpage

\end{document}